\documentclass[aos]{imsart}
\DeclareUnicodeCharacter{03B2}{$\beta$}
\RequirePackage{amsthm,amsmath,amsfonts,amssymb}
\usepackage{titlesec}
\usepackage{bibentry}
\usepackage{subcaption}
\usepackage{enumitem}
\usepackage{multirow}

\numberwithin{equation}{section} 
\usepackage[normalem]{ulem}
\RequirePackage[authoryear]{natbib}
\usepackage{hyperref}
\hypersetup{colorlinks,allcolors=blue}

\usepackage[algoruled,boxed,lined]{algorithm2e}

\usepackage{cleveref} 
\Crefname{proposition}{Proposition}{Propositions}
\Crefname{lemma}{Lemma}{Lemmas}
\Crefname{definition}{Definition}{Definitions}
\Crefname{theorem}{Theorem}{Theorems}
\Crefname{condition}{Condition}{Conditions}
\usepackage{autonum}
\usepackage{xr}
\RequirePackage{graphicx}

\newcommand{\Prob}{\mathbb{P}}

\newcommand{\E}{\mathbb{E}}
\newcommand{\rd}{\mathrm{d}}

\newcommand{\AVar}{\mathrm{AsyVar}}
\newcommand{\prs}[1]{\bigg( #1 \bigg)}
\newcommand{\prm}[1]{\bigg[ #1 \bigg]}
\newcommand{\prb}[1]{\bigg\{ #1 \bigg\}}

\newcommand{\ZILdis}{\text{ZIL}}

\newcommand{\Diam}{\text{diam}}

\def\be{\begin{equation}}
\def\ee{\end{equation}}

\startlocaldefs
\theoremstyle{plain}

\newtheorem{theorem}{Theorem}[section]

\newtheorem{corollary}{Corollary}[section]
\newtheorem{proposition}{Proposition}[section]
\newtheorem{condition}{Condition}

\def\be{\begin{equation}}
\def\ee{\end{equation}}
\def\bea{\begin{eqnarray}}
\def\eea{\end{eqnarray}}

\theoremstyle{remark}
\newtheorem{definition}{Definition}[section]

\endlocaldefs

\begin{document}

\begin{frontmatter}
\title{Versatile differentially private learning for general loss functions}
\runtitle{Versatile differentially private learning for general loss functions}

\begin{aug}
\author[A]{\fnms{Qi Long}~\snm{Lu}\ead[label=e1]{lu\_qilong@stu.pku.edu.cn}},
\author[B]{\fnms{Song Xi}~\snm{Chen}\ead[label=e2]{sxchen@tsinghua.edu.cn}}
\and
\author[C]{\fnms{Yu Mou}~\snm{Qiu}\ead[label=e3]{qiuyumou@math.pku.edu.cn}}

\address[A]{Guanghua School of Management, Peking University\printead[presep={,\ }]{e1}}
\address[B]{Department of Statistics and Data Science, Tsinghua University\printead[presep={,\ }]{e2}}
\address[C]{School of Mathematical Sciences and Center for Statistical Science, Peking University\printead[presep={,\ }]{e3}}

\end{aug}

\begin{abstract}
This paper aims to provide a versatile privacy-preserving release mechanism along with a unified approach for subsequent parameter estimation and statistical inference. We propose the ZIL privacy mechanism based on zero-inflated symmetric multivariate Laplace noise, which requires no prior specification of subsequent analysis tasks, allows for general loss functions under minimal conditions, imposes no limit on the number of analyses, and is adaptable to the increasing data volume in online scenarios. We derive the trade-off function for the proposed ZIL mechanism that characterizes its privacy protection level. Within the M-estimation framework, we propose a novel doubly random corrected loss (DRCL) for the ZIL mechanism, which provides consistent and asymptotic normal M-estimates for the parameters of the target population under differential privacy constraints. The proposed approach is easy to compute without numerical integration and differentiation for noisy data. It is applicable for a general class of loss functions, including non-smooth loss functions like check loss and hinge loss. Simulation studies, including logistic regression and quantile regression, are conducted to evaluate the performance of the proposed method. 
\end{abstract}

\begin{keyword}[class=MSC]
\kwd[Primary ]{62-11}
\kwd[; secondary ]{68P27}
\end{keyword}

\begin{keyword}
\kwd{Differential privacy}
\kwd{M-estimation}
\kwd{Symmetric multivariate Laplace distribution}
\kwd{Zero-inflated symmetric multivariate Laplace distribution}
\end{keyword}

\end{frontmatter}

\section{Introduction}
\label{sec:Intro}
Data privacy has become an increasing concern with the phenomenal growth in the amount of personal information stored in digital devices, such as health data, web search histories, and personal preferences \citep{erlingsson_rappor_2014,team_learning_2017,ding_collecting_2017}.
Analyzing data under privacy protection involves two roles: data providers and data analysts. The data provider stores the original data and releases privacy-preserved data or statistics to data analysts through a certain mechanism. The data analysts conduct analysis for various tasks based on the data or statistics released by the data provider \citep{Dwork06Foundation}.
Privacy-preserving mechanisms aim to protect the personal information in the original data while allowing data analysts to extract useful information from the outputs.

To quantify the privacy protection level of a data release mechanism, \cite{dwork2006our} proposed the concept of $(\epsilon,\delta)$-differential privacy ($(\epsilon,\delta)$-DP) that measures the similarity of the distributions of the outputs when the recording of an arbitrary sample is changed while all other samples are kept the same. 
Based on the DP framework, various privacy measures have been developed for different goals, for example, protecting explicit specification of the information in the data \citep{kifer2014pufferfish} and edge privacy for social network data \citep{nissim2007smooth}.
\cite{dong_gaussian_2022} linked differential privacy with the type I and type II error in hypothesis testing and proposed the concept of $f$-DP for a trade-off function $f$. 

Existing DP mechanisms can be broadly classified into three categories. One adds noise to data outputs, such as histograms \citep{lei_differentially_2011}, summary statistics or estimators \citep{avella-medina_privacy-preserving_2021}, which is called the sensitivity method \citep{dwork_calibrating_2006}.  
Another category introduces noises within  the computational procedure %
for a particular task.
The noisy stochastic gradient descent (Noisy-SGD) is a representative of this category, which adds noises to the SGD procedure of a specific loss function \citep{rajkumar_differentially_2012, bassily_private_2014}, and 
the objective function perturbation \citep{chaudhuri_differentially_2011} is another representative. 
For this category, the data analysts and the data provider need to communicate in each step of computation task, and the noise is added in each step by the data provider \citep{dong_gaussian_2022}.

These two categories of methods are not versatile as the data privacy protection is task-specific, and the data analysts can only conduct a limited number of analyses using perturbed statistics or SGD.   Specifically, we say a differential privacy mechanism is versatile if it is (i) applicable to general 
estimation tasks, (ii) applicable to any number of analyses, and (iii) adaptive to increasing data volume in online scenarios.

The third category of method 
adds noise to the original data \citep{warner1965,duchi_minimax_2018} or synthetic data \citep{zhang_differentially_2024} directly. It is more versatile than the first two categories as the noises are not tied to any analysis task and the corrupted data can be used repeatedly. 
Adding either the Laplace or Gaussian noise is a common practice \citep{wasserman_statistical_2010, dong_gaussian_2022}. 
However, the existing methods in this category require stronger conditions on the loss function for statistical inference. 
The subsequent analysis of the noisy data is closely related to the deconvolution method for the measurement error problems \citep{carroll_optimal_1988, fan_optimal_1991}. 
To remove the effects of the added noise in the estimation, the deconvoluted loss function can be formulated leading to a corrected loss
\citep{stefanski_unbiased_1989, wang_correctedloss_2012}. 
The forms of the corrected loss depend on the type of the added noise. 
For Gaussian noise, the corrected loss involves numerical integration because the inverse Fourier transform does not have a closed-form solution. 
For the component-wise independent Laplace noise, the corrected loss involves high-order differentiation of the underlying loss function. 
These suggest that applying both types of noise will impose strong conditions on the loss and bring inconvenience in the statistical estimation and inference. 

In recent years, there have been studies on statistical estimation for privacy-preserved data.
\cite{duchi_minimax_2018} considered estimating the mean, median, generalized linear model, and density function under the local differential privacy constraints. \cite{rohde_geometrizing_2020} focused on estimating parameters defined as a linear functional of the data distribution. \cite{cai_cost_2021} employed the Noisy-SGD to estimate linear regression coefficients in both low- and high-dimensional settings.  
All these works designed their data release mechanisms according to specific tasks, 
which inherently limits the ability to meet various analytical demands for a dataset while preserving privacy. This limitation makes the data privacy procedure less versatile.  
In addition to achieving the local differential privacy, we consider multivariate parameter estimation for the general M-estimation using a unified and versatile estimation method.

This paper aims to provide a versatile privacy-preserving mechanism that facilitates easier consistent parameter estimation and inference under the framework of M-estimation, without requiring the loss function to be smooth or numerical integration of the inverse Fourier transform of the loss. As a result, the procedure would permit a wide range of inference tasks with even non-smooth loss functions. 
To achieve these goals, we consider
the zero-inflated symmetric multivariate Laplace (ZIL) distribution as the noise distribution. 
The ZIL noise distribution is designed to 
simplify the subsequent parameter estimation and statistical inference avoiding the need to compute derivatives of the loss function or perform integrations in the deconvoluted loss. 

The DP properties of the ZIL mechanism are developed under the $f$-DP framework by deriving the trade-off function and their asymptotic limits as the data dimension diverges. 
The connection between the ZIL trade-off function and the $(\epsilon, \delta)$-DP 
criteria is derived which allows interpretation of the DP level via the $(\epsilon, \delta)$-DP criteria and provides a guideline for selecting the noise level and the zero proportion of the ZIL distribution under a given privacy budget.

To facilitate consistent M-estimation with data released under the ZIL mechanism, we propose a doubly random (DR) procedure that additionally adds symmetric multivariate Laplace (SL) noises to the output of ZIL mechanism to construct a corrected loss function which is unbiased to the underlying expected loss. The proposed method is versatile for a general class of M-estimation that does not require differentiation or numerical Fourier integration and is free of tuning parameters as would be for the case of the deconvolution density estimation. 
It works for non-smooth loss functions including quantile regression, classification using the hinge loss for support vector machines, and neural network models using the ReLU activation function.  

We show that the proposed double random corrected loss (DRCL) estimator is consistent and asymptotically normal. The variance of the DRCL estimator is obtained, which can be easily estimated for inference purposes. 
Compared with the estimation procedures with the well-known Gaussian and Laplace mechanisms that add independent normal and Laplace errors respectively, the proposed DRCL procedure is much simpler avoiding integration (Gaussian noise case) and differentiation (Laplace noise case), and works for more general loss functions without requiring their being smooth. 
%
If the data analysts' tasks are constrained to second-order smooth loss functions,  we further propose a smoothed doubly random corrected loss (sDRCL) that utilizes the smoothness of the loss function for parameter estimation. 
We show that the sDRCL estimator achieves a smaller asymptotic variance than the DRCL estimator in the linear regression setting.

The paper is organized as follows. Section \ref{sec:background} reviews the necessary concepts and properties of differential privacy and $f$-DP, and introduces a related concept, attribute differential privacy (ADP), that measures the privacy protection level for each attribute of the data. Section \ref{sec:privacy-guarantee} %
describes the proposed ZIL mechanism and derives its trade-off function and the associated properties. Section \ref{sec:M-estimation} proposed the doubly random corrected loss method for parameter estimation under the ZIL mechanism. Section \ref{sec:DRCL} establishes the theoretical results for the proposed DRCL estimator. Section \ref{sec:alternatives} provides an alternative estimator by smoothed doubly random corrected loss for second-order smooth loss functions and analyzes the efficiency of the DRCL and sDRCL estimators. Section \ref{sec:simulation} conducts simulation studies to verify the theoretical findings. All the technical proofs and additional numerical results are relegated to the supplementary material (SM).


\section{Background on Differential Privacy}\label{sec:background}

Let $\mathcal{X} \subset \mathbb{R}^{d}$ denote the space of the $d$-dimensional data $X$ from an individual, and $\mathcal{X}^{n}$ denote the space of a dataset $\mathbf{X}=\{X_{1},\ldots,X_{n}\}$ containing $n$ individuals, where $X_i = (X_{i1}, \ldots, X_{id})^{\top}$, and the superscript $n$ denotes $n$ Cartesian products of $\mathcal{X}$. Let $\mathcal{X}_j$ denote the space of each component of $X$ for $j = 1, \dots, d$. A differential privacy (DP) mechanism \( \text{Mech}(\cdot) \) is a randomized algorithm that releases some (randomized) statistics or noisy data of the input dataset. It is a randomized mapping defined on the space of datasets $\mathcal{X}^{n}$ to some abstract space $\mathcal{R}$. Randomized mapping means the data release mechanism would add noises into the input dataset to preserve privacy. For two datasets $\mathbf{X}=\{X_{1},\ldots,X_{n}\}$ and $\mathbf{X}'=\{X'_{1},\ldots,X'_{n}\}$, define 
\be\label{eq:diff-global}
\Delta_{\text{I}}(\mathbf{X}, \mathbf{X}') = |\{i: X_{i} \neq X'_{i}, 1 \leq i \leq n \}|\ee
be the number of individuals with different values for their data records, where $|\mathcal{A}|$ denotes the cardinality of a set $\mathcal{A}$. Here, the subscript ``I'' denotes the individual-level difference to differentiate from the attribute-level difference in \eqref{eq:diff-attribute} where the subscript ``A" will be used.
A DP mechanism would make the distributions of $\text{Mech}(\mathbf{X})$ and $\text{Mech}(\mathbf{X}')$ being similar for any pairs of neighboring datasets $\mathbf{X}$ and $\mathbf{X}'$ with only one individual having different records so that the information of $X_i$ is preserved for all $i = 1, \ldots, n$ under this mechanism.  

     \begin{definition}[$(\varepsilon,\delta)$-DP \citep{dwork_calibrating_2006}]\label{def:Dwork}
    
    For any  non-negative $\varepsilon$ and $\delta$, a randomized algorithm $\text{Mech}\ :\mathcal{X}^{n}\to\mathcal{R}$ is $(\varepsilon,\delta)$-differentially private if for every pair of data sets $\mathbf{X},\mathbf{X}'\in\mathcal{X}^{n}$ with $\Delta_{\text{I}}(\mathbf{X}, \mathbf{X}') = 1$ and every measurable set
    $S\subseteq\mathcal{R}$, 
    $$\mathbb{P}(\text{Mech}(\mathbf{X})\in S)\leqslant e^{\varepsilon}\cdot\mathbb{P}(\text{Mech}(\mathbf{X}')\in S)+\delta,$$
    where the probability measure $\mathbb{P}$ is conditioned on the data sets $\mathbf{X},\mathbf{X}'$ and is induced by the randomness of $\text{Mech}(\cdot)$ only. 
    \end{definition} 

As differential privacy is measured by the similarity between the conditional distributions of $\text{Mech}(\mathbf{X})$ and $\text{Mech}(\mathbf{X}')$ given $\mathbf{X}$ and $\mathbf{X}'$, it can be characterized from the perspective of testing the hypotheses \citep{dong_gaussian_2022}:
\be \label{eq:hypothesis0}
H_{0}: \text{the original dataset is}\ \mathbf{X}\ \ \ \ \text{vs.}\ \ \ \ H_{1}: \text{the original dataset is}\ \mathbf{X}'
\ee
based on the released results from the mechanism $\text{Mech}$.
Let $P$ and $Q$ be two generic notations for hypotheses testing, representing the distributions under the null and alternative hypotheses,
respectively. 
The probabilities of type I and type II errors of a test function $\phi$ are, respectively,
$$\alpha_{\phi}:=\mathbb{E}_{P}[\phi] \mbox{ \ and \ } \beta_{\phi}:=1-\mathbb{E}_{Q}[\phi],$$
where $\mathbb{E}_{P}[\phi]:=\int\phi(x){\rm d}P$ and $\mathbb{E}_{Q}[\phi]:=\int\phi(x){\rm d}Q$.
The trade-off function $T(P, Q):[0,1]\rightarrow[0,1]$ for distinguishing $P$ and $Q$ is defined as 
    \be \label{eq:tradeoff}
    T(P, Q)(\alpha)=\inf_{\phi}\{\beta_{\phi}:\alpha_{\phi}\leqslant\alpha\}
	\ee	where the infimum is taken over all measurable test functions.
For the hypotheses in \eqref{eq:hypothesis0}, to simplify notations, we also use $\text{Mech}(\mathbf{X})$ and $\text{Mech}(\mathbf{X}')$ to denote the conditional distributions of the released results given $\mathbf{X}$ and $\mathbf{X}'$, respectively, when there is no confusion.
The trade-off function $T(\text{Mech}(\mathbf{X}), \text{Mech}(\mathbf{X}'))(\alpha)$ fully characterizes the optimal test for distinguishing the original data being $\mathbf{X}$ or $\mathbf{X}'$. The definition of $f$-DP is based on a trade-off function $f$, presented in the following.

\begin{definition}[$f$-DP \citep{dong_gaussian_2022}]\label{f-DP}
    Let $f$ be a trade-off function. A mechanism $\text{Mech}$ is said to be $f$-differentially private ($f$-DP) if for all $\alpha\in[0,1]$ $$T(\text{Mech}(\mathbf{X}),\text{Mech}(\mathbf{X}'))(\alpha)\geqslant f(\alpha)$$
for all neighboring datasets $\mathbf{X}, \mathbf{X}'\in\mathcal{X}^{n}$ with $\Delta_{\text{I}}(\mathbf{X}, \mathbf{X}') = 1$. 
\end{definition}

In \Cref{f-DP}, the measure of privacy protection is reflected by the function $f$. For each $\alpha \in (0,1)$, if the probability of the analyst's type I error for distinguishing two adjacent datasets is less than $\alpha$, then the probability of the type II error must be above $f(\alpha)$. 
For two trade-off functions $f_1$ and $f_2$, if $f_1(\alpha) \geqslant f_2(\alpha)$ for all $\alpha \in [0,1]$, higher level of privacy is protected by an $f_1$-DP mechanism. However, statistical inference would be more difficult as larger noises need to be added to this mechanism. 

Let $G_{\mu}(\alpha):=\Phi(\Phi^{-1}(1-\alpha)-\mu)$ 
for $\alpha\in[0,1]$, where  $\Phi(\cdot)$ and $\Phi^{-1}(\cdot)$ denote the cumulative distribution function and quantile function of the standard normal distribution, respectively. A special case of $f$-DP is the $\mu$-GDP (Gaussian differential privacy) with $f = G_{\mu}$, built upon the standard Gaussian noise. As $G_{\mu}(\alpha)$ monotonically decreases as $\mu$ increases for a given $\alpha$, a smaller $\mu$ indicates higher privacy protection level of a $\mu$-GDP mechanism. 

Let ${\rm Proc}:\mathcal{R}\to\mathcal{Z}$ denote a randomized algorithm that maps the released result $\text{Mech}(\mathbf{X})\in\mathcal{R}$ of an $f$-DP mechanism $\text{Mech}$ to some space $\mathcal{Z}$, yielding a new mechanism denoted by ${\rm Proc}\circ \text{Mech}$. 
\cite{dong_gaussian_2022} showed the following two propositions of an $f$-DP mechanism. 

\begin{proposition}[\cite{wasserman_statistical_2010, dong_gaussian_2022}]\label{thm:relation}
A mechanism $\text{Mech}$ is $(\varepsilon,\delta)$-DP if and only if the $\text{Mech}$ is $f_{\varepsilon,\delta}$-DP where $f_{\varepsilon,\delta}(\alpha):=\max\{0,\ 1-\delta-e^{\varepsilon}\alpha,\ e^{-\varepsilon}(1-\delta-\alpha)\}$ for $\alpha\in[0,1]$.   
\end{proposition}

\begin{proposition}
	[\cite{dong_gaussian_2022}]\label{thm:post-processing}
	If a mechanism $\text{Mech}$ is $f$-DP, then its post-processing ${\rm Proc}\circ \text{Mech}$ is also $f$-DP.
\end{proposition}

 \Cref{thm:relation} shows the equivalence between $f$-DP and $(\varepsilon,\delta)$-DP. \Cref{thm:post-processing} shows that post-processing a mechanism does not compromise the privacy guarantees already provided by the mechanism. This property ensures that the privacy protection level of an estimator computed from the output of an $f$-DP mechanism is preserved.

The measures of differential privacy in  \Cref{def:Dwork,f-DP} depend on the definition of the distance for neighboring datasets. Under the distance $\Delta_{\text{I}}(\mathbf{X}, \mathbf{X}')$ in \eqref{eq:diff-global}, if an analyst cannot distinguish any pair of datasets $\mathbf{X}, \mathbf{X}'\in\mathcal{X}^{n}$ with $\Delta_{\text{I}}(\mathbf{X}, \mathbf{X}') = 1$, the analyst would not know whether any particular individual is part of the original dataset based on the output of the privacy mechanism.
However, in some scenarios, it is unnecessary to protect the privacy of all variables of an individual as a whole \citep{kifer2014pufferfish}. For example, in survey sampling of yearly income, we may not need to preserve the information of which individual is sampled but only to ensure that each attribute of each individual cannot be inferred from the mechanism's output. We refer to this relaxed version of DP as attribute-level DP (ADP). 
Let
\be\label{eq:diff-attribute}
\Delta_{\text{A}}(\mathbf{X}, \mathbf{X}') = |\{(i, j) : X_{ij} \neq X'_{ij}, 1 \leq i \leq n, 1 \leq j \leq d \}|
\ee
be the attribute-level distance between two datasets $\mathbf{X}$ and $\mathbf{X}'$. In the following, we formally define ADP under the attribute-level distance $\Delta_{\text{A}}(\mathbf{X}, \mathbf{X}')$, which relaxes the $f$-DP in \Cref{f-DP} under the global distance $\Delta_{\text{I}}(\mathbf{X}, \mathbf{X}')$.

\begin{definition}[Attribute differential privacy]\label{def:attribute_level_dp}
Let $f$ be a trade-off function. A mechanism $\text{Mech}$ is said to be $f$-attribute differentially private ($f$-ADP) if for all $\alpha\in[0,1]$ $$T(\text{Mech}(\mathbf{X}),\text{Mech}(\mathbf{X}'))(\alpha)\geqslant f(\alpha)$$
for all neighboring datasets $\mathbf{X}, \mathbf{X}'\in\mathcal{X}^{n}$ with $\Delta_{\text{A}}(\mathbf{X}, \mathbf{X}') = 1$. 
Furthermore, a mechanism is said to be $(\varepsilon,\delta)$-ADP if it is $f_{\varepsilon,\delta}$-ADP.
\end{definition}

An \(f\)-ADP mechanism is characterized by the hypotheses to distinguish two neighboring datasets which only differ in one attribute of one individual under the attribute-level distance $\Delta_{\text{A}}(\mathbf{X}, \mathbf{X}')$. 
Compared to $f$-DP which prevents data analysts from distinguishing any difference in each individual based on the output of a mechanism, \(f\)-ADP prevents data analysts from distinguishing any difference in each attribute of each individual. It also requires adding perturbation to every attribute in the data.
Since \(f\)-ADP is simply \(f\)-DP based on a different definition of neighboring datasets, it inherits all the properties of \(f\)-DP. Note that the edge differential privacy for social network data \citep{nissim2007smooth, chang_edge_2024}, which aims to protect the information of whether each edge exists or not in a network, is a special form of ADP. 
When the original data for each individual is one-dimensional, \(f\)-ADP is equivalent to \(f\)-DP. However, when the dimension $d > 1$, \(f\)-ADP is a more relaxed measure of differential privacy than \(f\)-DP, and hence allows for higher data utility.

In the following sections, we will construct a differentially private mechanism that releases noisy data with a novel noise distribution, which is applicable for general analysis tasks chosen by the analyst under both frameworks of $f$-DP and $f$-ADP. The innovations of the proposed procedure for differentially private learning are to achieve versatile estimation and statistical inference for a variety of loss functions under a guaranteed privacy level by carefully designing the noise distribution and a new denoising approach.

\section{Zero-Inflated Multivariate Laplace Mechanism and its Privacy Guarantee}\label{sec:privacy-guarantee}

We outline the new privacy protection mechanism that adds the zero-inflated symmetric multivariate Laplace (\text{ZIL}) noises. 
Adding noise directly to the data makes the ZIL mechanism versatile as it requires no prior specification of subsequent analysis tasks, imposes no limit on the number of analyses, and is adaptable to the increasing data volume in online scenarios. 
The extent of the differential privacy guarantee will be studied as well.

\subsection{Zero-Inflated Symmetric Multivariate Laplace Distribution.}\label{sec:intro_ZIL} 

We first introduce the symmetric multivariate Laplace (SL) distribution $\mathcal{SL}_{d}(\Sigma)$ of dimension $d$ with covariance matrix $\Sigma$ defined via its characteristic function $\Psi_{\Sigma}^{\mathcal{SL}}(t) = (1 + t^{\top}\Sigma t / 2)^{-1}$, which reduces to the Laplace distribution when $d=1$. 
However, for $d>1$, $\mathcal{SL}_{d}(I_{d})$ does not represent the distribution of $d$ independent Laplace random variables, where $I_{d}$ denotes the $d$-dimensional identity matrix.
Let $f_{\Sigma}^{\mathcal{SL}}(x)$ denote the density of $\mathcal{SL}_{d}(\Sigma)$.
A random vector $S \sim \mathcal{SL}_{d}(\Sigma)$ following the symmetric multivariate Laplace distribution can be generated by $S = \sqrt{W} X$, where $W$ and $X$ are independent, $W$ follows the exponential distribution $\mbox{Exp}(1)$, and $X\sim N(0,\Sigma)$ \citep{kotz_laplace_2001}.  

For $\delta\in(0,1)$ and a covariance matrix $\Sigma$, we define the zero-inflated symmetric multivariate Laplace (ZIL) distribution $\text{ZIL}(\delta,\Sigma)$ as a mixture distribution of the point mass at $0$ and the symmetric multivariate Laplace distribution $\mathcal{SL}_{d}(\Sigma)$. Let $S \sim \mathcal{SL}_{d}(\Sigma)$ and $\xi$ follow the binary distribution $\mbox{Bern}(1, \delta)$. The ZIL random variable $Z \sim \text{ZIL}(\delta,\Sigma)$ is generated by $Z = S \mathbf{1}(\xi = 0)$, which equals to 0 with probability $\delta$ and equals to $S$ with probability $1 - \delta$, where $\mathbf{1}( \cdot )$ denotes the indicator function. 
The characteristic function $\Psi_{\delta,\Sigma}^{\text{ZIL}}(t)$ of $\text{ZIL}(\delta,\Sigma)$ is 
$$\Psi_{\delta,\Sigma}^{\text{ZIL}}(t) = \delta + (1-\delta) \Psi_{\Sigma}^{\mathcal{SL}}(t) = (1 + \delta t^{\top}\Sigma t / 2)(1 + t^{\top}\Sigma t / 2)^{-1}.$$

\subsection{ZIL Mechanism and Privacy Guarantee.}\label{sec:privacy_level}

We introduce the ZIL mechanism for differential privacy and derive its trade-off function and the associated properties as follows. 

\begin{definition}[ZIL mechanism]\label{def:ZIL}
    Suppose that $\{X_i\}_{i=1}^{n}$ are $d$-dimensional random vectors {with a compact support $\mathcal{X}$}. 
    The ZIL mechanism, which is a randomized algorithm, is 
    \begin{equation}\label{eq:LapM}
        \text{Mech}_{\text{ZIL}}(\{X_i\}_{i=1}^{n};\delta,\lambda)=\{X_{i}+Z_{i}\}_{i=1}^{n},
    \end{equation}
    where $Z_{1},\ldots,Z_{n}\overset{\text{i.i.d.}}{\sim}\text{ZIL}(\delta,\lambda^{2}I_{d})$ for a $\lambda>0 $ and $\delta\in(0,1)$.
    \end{definition} 

{Similarly, the symmetric multivariate Laplace (SL) mechanism adds noises from the SL distribution $\mathcal{SL}_{d}(\Sigma)$ to the original data. To derive the trade-off function for the ZIL mechanism, we first introduce that for the SL mechanism with the identity covariance matrix. As the SL distribution with identity covariance is spherically symmetric, testing the hypotheses in \eqref{eq:hypothesis0} under the distribution $\mathcal{SL}_{d}(I_d)$ is equivalent to testing}
\begin{equation}\label{eq:hypothesis2}
    H_{0}: P = \mathbf{0}_{d} + \mathcal{SL}_{d}(I_{d}) \quad \text{versus} \quad H_{1}: Q = (c,0,\ldots,0)^{\top} + \mathcal{SL}_{d}(I_{d})
\end{equation}
for some constant $c$. 
Let
\begin{equation}\label{eq:def-T}
    T_{d,c}(\alpha):=T(\mathcal{SL}_{d}(I_{d}), (c, 0, \ldots, 0)^{\top} + \mathcal{SL}_{d}(I_{d}))(\alpha) 
\end{equation} 
{be the trade-off function for the hypotheses in \eqref{eq:hypothesis2}. It is shown in the SM that $T_{d,c}(\alpha)$ is the trade-off function for the SL mechanism. Let}
\begin{align}
T_{d,c,\delta}(\alpha) = 
\begin{cases} 
    0, & \text{if } \alpha>1-\delta ,\\
    (1-\delta)T_{d,c}\big(\frac{\alpha}{1-\delta}\big), & \text{if } 0\leqslant\alpha\leqslant 1-\delta\label{eq:shrink-dp}
\end{cases}
\end{align} 
for $\delta\in(0,1)$. 
{ As $T_{d,c}(\alpha)$ is a trade-off function, $T_{d, c, \delta}(\alpha)$ is also a trade-off function.}
Define the diameter of a set $A$ as $\mathrm{diam}(A)=\sup_{x,x'\in A}\|x-x'\|_{2}$. 
The following theorem shows the privacy protection of the ZIL mechanism. 

\begin{theorem}
\label{thm:privacy-of-ZIL}
 The ZIL mechanism in \Cref{def:ZIL} is
 \(T_{d,c_{A},\delta}\)-ADP and \(T_{d,c_{I},\delta}\)-DP for $c_{A} = \max_{j\in\{1,\ldots,d\}}\mathrm{diam}(\mathcal{X}_{j})/\lambda$ and $c_{I} = \mathrm{diam}(\mathcal{X})/\lambda$, respectively. 
\end{theorem}

{The theorem shows that the ZIL mechanism is a special form of $f$-DP and $f$-ADP, which suggests that the privacy level of the ZIL mechanism depends on $T_{d,c}(\alpha)$ in} \eqref{eq:def-T}.  The following proposition presents the properties of \( T_{d,c}(\alpha) \) and \( T_{d,c,\delta}(\alpha) \), including the monotonicity with respect to \( c \), \( d \), and \( \alpha \), respectively.

\begin{proposition}\label{lemma:compare}
    (i) For any $d$, if $c_{1}\geqslant c_{2}>0$,  $T_{d,c_{1}}(\alpha)\leqslant T_{d,c_{2}}(\alpha)$ for $\alpha\in[0,1]$.

    (ii) For any $c>0$, if $d_{1}\geqslant d_{2}$, %
    then $T_{d_{1},c}(\alpha)\leqslant T_{d_{2},c}(\alpha)$ for $\alpha\in[0,1]$.

    (iii) For any $d$ and $c>0$, $T_{d,c}(\alpha)$ is 
    decreasing with respect to $\alpha$, convex and continuous. \\ 
    \, \, Furthermore, $T_{d,c}(\alpha)\leqslant 1-\alpha$ for $\alpha\in[0,1]$ and is symmetric about the 45-degree line such that $T_{d,c}(\alpha) = T_{d,c}^{-1}(\alpha)$ for $\alpha\in[0,1]$, where $T_{d,c}^{-1}(\alpha) = \inf\{t\in[0,1]: T_{d,c}(t)\leqslant \alpha\}$.

    (iv) Part (i)--(iii) also holds for $T_{d, c, \delta}(\alpha)$.
\end{proposition}

Note that $T_{d,c}(\alpha)$ is the type II error of the most powerful test for the hypotheses in \eqref{eq:hypothesis2}
at the significance level $\alpha$. 
Given a data point \( s = (s_{1}, \ldots, s_{d})^{\top} \), 
we need to determine whether it originates from the distribution \( P \) or \( Q \) in \eqref{eq:hypothesis2}. Recall that $f_{I_d}^{\mathcal{SL}}(x)$ is the density of $\mathcal{SL}_{d}(I_d)$. According to the Neyman-Pearson Lemma, the most powerful test $\phi^{ump}$ at the significance level \( \alpha \) is 
\[
\phi^{ump}(s; k) = 
\begin{cases}
  1, & \text{if } f^{\mathcal{SL}}_{I_{d}}(s-(c,0,\ldots,0)^{\top})>k f^{\mathcal{SL}}_{I_{d}}(s),\\
  0, & \text{if } f^{\mathcal{SL}}_{I_{d}}(s-(c,0,\ldots,0)^{\top})\leqslant k f^{\mathcal{SL}}_{I_{d}}(s),
\end{cases}
\]
where $k$ satisfies %
that $\E_{S\sim P}\phi^{ump}(S; k) = \alpha$.  
The probabilities of type I and type II errors are  \( a(k) = \mathbb{E}_{S \sim P}\{\phi^{\text{ump}}(S; k)\} = \alpha \) and \( b(k) = 1 - \mathbb{E}_{S \sim Q}\{\phi^{\text{ump}}(S; k)\} \), respectively. 
Thus, the trade-off function \( T_{d,c}(\alpha)=T(P,Q)(\alpha) = b(a^{-1}(\alpha)) \). 
Deriving a closed-form expression for the trade-off function for  \( d > 1 \) is challenging. To appreciate this, we note that  the density \( f_{\Sigma}^{\mathcal{SL}} \) of \( \mathcal{SL}_{d}(\Sigma) \), according to \cite{kotz_laplace_2001}, %
\begin{equation}
f_{\Sigma}^{\mathcal{SL}}(x) = 2 (2\pi)^{-\frac{d}{2}} |\Sigma|^{-\frac{1}{2}} (x^{\top}\Sigma^{-1}x / 2)^{\frac{v}{2}} K_{v}\{ (2 x^{\top}\Sigma^{-1}x)^{1/2} \},
\end{equation}
where $v=(2-d)/2$ and $K_{v}(u)$ is the modified Bessel function of the third kind. 
The complexity of \( f^{\mathcal{SL}}_{I_{d}} \) makes deriving a closed-form expression for \( T_{d,c}(\alpha)\) challenging due to the involvement of the modified Bessel function. 
Simulation is a viable approach to attain values for $T_{d,c}$. It generates samples from the distributions \( P \) and \( Q \) in \eqref{eq:hypothesis2}, and simulates \( a(k) \) and \( b(k) \), leading to the empirical trade-off curve \( \{(a(k), b(k)):k\in[0,\infty)\} \).

However, for \( d = 1 \),  a tangible expression is available for $T_{1,c}$. As \( \mathcal{SL}_{1}(I_1) \) is the Laplace distribution \( L(\sqrt{1/2}) \), it can be proved that $T_{1,c}(\alpha)=F_{\text{Lap}}(F_{\text{Lap}}^{-1}(1-\alpha)-\sqrt{2}c)$, where $F_{\text{Lap}}$ is the cumulative distribution function of the standard Laplace distribution $L(1)$. In this case, according to \cite{dwork_calibrating_2006}, \( T_{1,c}(\alpha) \geqslant f_{\sqrt{2}c/\lambda,0}(\alpha) \) for \( \alpha \in [0,1] \), where $f_{\sqrt{2}c/\lambda,0}(\alpha)$ is given in \Cref{thm:relation}.


The monotonicity of $T_{d,c}$ and $T_{d, c, \delta}$ with respect to $d$ leads us to consider the asymptotic trade-off functions $T_{d, c}(\alpha)$ and $T_{d, c, \delta}(\alpha)$ for \( d \to \infty \), which turns out to be trade-off functions themselves as shown below.
%
Recall that a SL random vector $S = (S_1, \ldots, S_d)^\top \sim \mathcal{SL}_{d}(I_{d})$ can be generated by $S = \sqrt{W}X$, where $W \sim \mbox{Exp}(1)$, $X=(X_{1},\ldots,X_{d})^{\top} \sim N(0, I_d)$ and $W$ is independent of $X$. 
{It is noted that the most powerful test for the hypotheses in \eqref{eq:hypothesis2} can be written as the conditional density ratio of $S_1$ given $S_2, \ldots, S_d$, and }
\begin{equation} 
(d-1)^{-1}(S_{2}^{2} + \ldots + S_{d}^{2}) \overset{a.s.}{\to} W\text{ as }d\to\infty
\end{equation} 
according to the law of large numbers.  
These inspire us to consider testing the hypotheses %
\begin{equation}\label{eq:hypothesis3} 
H_{0}: P = (0 + \sqrt{W}X_{1},W) \quad \text{versus} \quad H_{1}: Q = (c + \sqrt{W}X_{1},W). 
\end{equation} 
Let $F_{c}(x)=\int_{0}^{\infty}\Phi( c^{-1} w^{1/2} x + c (4 w)^{-1/2})e^{-w}\,\mathrm{d}w$ be the cumulative distribution function of $c X_1 W^{-1/2} - c^{2}(2W)^{-1}$. The following proposition derives the trade-off function for the hypotheses \eqref{eq:hypothesis3}.

\begin{proposition}\label{lemma:closed-form}
(i) For the hypotheses in \eqref{eq:hypothesis3} with $c > 0$, its trade-off function is
\begin{equation}
    \beta_{c}(\alpha)=\int_{0}^{\infty}\Phi\bigg\{ \frac{\sqrt{w}F_{c}^{-1}(1-\alpha)}{c}-\frac{c}{2\sqrt{w}} \bigg\} e^{-w}\rd w,
\end{equation}
where $\beta_{c}(0)=1$, $\beta_{c}(1)=0$ and $\beta_{c}(\alpha)>0$ for $\alpha\in[0,1)$.

(ii) Furthermore, $\beta_{c}(\alpha)$ is symmetric about the 45-degree line, strictly decreasing, convex, continuous, and $\beta_{c}(\alpha)\leqslant 1-\alpha$ for $\alpha\in[0,1]$. 
\end{proposition}

Proposition \ref{lemma:closed-form} shows that $\beta_{c}(\alpha)$ is the probability of type II error of the most powerful test for the hypotheses in \eqref{eq:hypothesis3} with $\alpha$ significance level. Let $h_c(\alpha) = c^{-1}F_{c}^{-1}(1-\alpha)$. It can be shown that 
\begin{equation}
    \beta_{c}(\alpha) = \bigg[ 1+\bigg\{\frac{\sqrt{2}}{h_c(\alpha) + \sqrt{2 + h_c^2(\alpha)}} \bigg\}^{2} \bigg]^{-1} \exp\prb{-\frac{c}{h_{c}(\alpha)+\sqrt{2+h_{c}^{2}(\alpha)}}}, 
\end{equation}
which suggests that $\beta_c(\alpha)$ can be readily computed after obtaining $F_c(1-\alpha)$. 
Let 
\begin{align}
\beta_{c,\delta}(\alpha) = 
\begin{cases} 
    0, & \text{if } \alpha > 1-\delta, \\
    (1-\delta)\beta_{c}\big(\frac{\alpha}{1-\delta}\big), & \text{if } 0 \leq \alpha \leq 1-\delta. 
\end{cases}
\end{align}
As $\beta_c(\alpha)$ is a trade-off function, $\beta_{c,\delta}(\alpha)$ is also a trade-off function. The following theorem shows that $\beta_{c}(\alpha)$ and $\beta_{c,\delta}(\alpha)$ are the asymptotic limits of $T_{d,c}(\alpha)$ and $T_{d, c, \delta}(\alpha)$ as $d \to \infty$, and are the lower bounds for all $T_{d,c}(\alpha)$ and $T_{d,c, \delta}(\alpha)$, respectively.  

\begin{theorem}\label{thm:asymptotic trade-off}
(i) For any $c>0$ and $\delta \in (0, 1)$,  
    $\lim_{d\to\infty}\sup_{\alpha\in[0,1]}|T_{d,c}(\alpha)-\beta_{c}(\alpha)|=0$ and $\lim_{d\to\infty}\sup_{\alpha\in[0,1]}|T_{d, c, \delta}(\alpha)-\beta_{c, \delta}(\alpha)|=0$.
    
(ii) For every $d\geqslant1$, $c>0$ and $\delta \in (0, 1)$, $T_{d,c}(\alpha)\geqslant \beta_{c}(\alpha)$ and $T_{d, c, \delta}(\alpha)\geqslant \beta_{c, \delta}(\alpha)$ for every $\alpha\in[0,1]$.
\end{theorem}  

{Note that $\beta_{c, \delta}(\alpha)$ is obtained by substituting $T_{d,c}(\alpha)$ in \eqref{eq:shrink-dp} by  its limit $\beta_{c}(\alpha)$. Since $\beta_{c}(\alpha)$ is an approximation of $T_{d,c}(\alpha)$, $\beta_{c, \delta}(\alpha)$ is an approximation of $T_{d,c,\delta}(\alpha)$ for the ZIL mechanism when $d$ is large.
\Cref{thm:asymptotic trade-off} (ii) suggests that we can use $\beta_{c, \delta}$ as a lower bound for $T_{d, c, \delta}$, while \Cref{thm:asymptotic trade-off} (i) ensures that the lower bound is tight 
when $d$ is large, meaning that $\beta_{c, \delta}$ can be used to describe the 
privacy protection level of the ZIL mechanism. Figure \ref{fig:1} (a) presents $T_{d, 0.5, 0.05}$ curves for selected values of $d$ and $\beta_{0.5, 0.05}$, which shows that $\beta_{0.5, 0.05}$ can well approximate $T_{d, 0.5, 0.05}$ for $d$ as small as 4.}

With $T_{d,c}$, we can now explain the rationale behind introducing attribute-level differential privacy (ADP) in relation to the individual DP.
It is noted that 
if $c(d)$ as a function of $d$ diverges such that $\lim_{d \to \infty} c(d) = \infty$, then $\lim_{d \to \infty} T_{d, c(d)}(\alpha) \leqslant \lim_{d \to \infty} T_{1,c(d)}(\alpha) = 0$ for $\alpha \in (0,1)$, which implies that the differential privacy guarantee tends to disappear if $c(d) \to \infty$.  
In \Cref{thm:privacy-of-ZIL}, $c_{A} = \max_{j \in \{1, \dots, d\}} \mathrm{diam}(\mathcal{X}_j) / \lambda$ for attribute differential privacy. If the range of each component $\mathrm{diam}(\mathcal{X}_j)$ is the same, then as $d$ increases, $c_{A}$ remains unchanged, which prevents 
the degradation of the attribute differential privacy (ADP) for any fixed $\lambda$. {In contrast, for the individual level differential privacy, $c_{I} = \mathrm{diam}(\mathcal{X}) / \lambda$. 
For example, $\mathcal{X}_{j}=[-M,M]$, the $\Diam(\mathcal{X})=\sum_{j=1}^{d}=4dM^{2}$ will diverge to infinity as $d \to \infty$, implying the individual-level DP  cannot be protected as more variables are collected from each individual if the noise variance (related to $\lambda$) is unchanged with respect to the dimension.}

{The $\beta_{c}$-DP ($\beta_{c}$-ADP) and $\beta_{c,\delta}$-DP ($\beta_{c,\delta}$-ADP) are special forms of the $f$-DP ($f$-ADP) defined in \Cref{f-DP} (\Cref{def:attribute_level_dp}). 
The following theorem provides the level of $(\varepsilon,\delta)$-DP ($(\varepsilon,\delta)$-ADP) achieved by the $\beta_{c,\delta}$-DP ($\beta_{c,\delta}$-ADP).} 

\begin{theorem}
    \label{thm:profile}
     (i) A mechanism is $\beta_{c}$-DP if and only if it is $(\varepsilon,\delta_{c}(\varepsilon))$-DP for all $\varepsilon\geqslant0$ where
    \begin{equation}\label{eq:profile}
        \delta_{c}(\varepsilon)=1-e^{\varepsilon}(1-F_{c}(\varepsilon))-\prm{1+\prb{\frac{\sqrt{2}}{(\varepsilon/c) + \sqrt{2+(\varepsilon/c)^{2}}}}^{2}}^{-1}\exp\prb{\frac{-c}{(\varepsilon/c)+\sqrt{2+(\varepsilon/c)^2}}}.  
    \end{equation}
Furthermore, a mechanism is $\beta_{c,\delta}$-DP if and only if it is $(\varepsilon,\tilde{\delta}_{c,\delta}(\varepsilon))$-DP for all $\varepsilon\geqslant0$, where $\tilde{\delta}_{c,\delta}(\varepsilon)=1-(1-\delta)\cdot(1-\delta_{c}(\varepsilon))$.

(ii) The same correspondence holds between $\beta_{c}$-ADP and $(\varepsilon,\delta_{c}(\varepsilon))$-ADP, and between $\beta_{c,\delta}$-ADP and $(\varepsilon,\tilde{\delta}_{c,\delta}(\varepsilon))$-ADP, respectively. 
\end{theorem}  

\Cref{thm:profile} indicates that there {are families of $f_{\varepsilon,\delta_{c}(\varepsilon)}$ and $f_{\varepsilon, \tilde{\delta}_{c, \delta}(\varepsilon)}$ envelopes for $\beta_{c}$ and $\beta_{c, \delta}$, respectively, as shown in \Cref{fig:1}(a).
From \Cref{thm:privacy-of-ZIL} and \Cref{thm:asymptotic trade-off}, \(\mathrm{Mech}_{\text{ZIL}}(\cdot; \delta, \lambda)\) satisfies \(\beta_{c_{A}, \delta}\)-ADP and \(\beta_{c_{I}, \delta}\)-DP, where \(c_{A} = \max_{j \in \{1, \ldots, d\}} \mathrm{diam}(\mathcal{X}_{j}) / \lambda\) and \(c_{I} = \mathrm{diam}(\mathcal{X}) / \lambda\). 

\begin{figure}[tp]
    \centering
    \begin{subfigure}[b]{0.45\textwidth}
        \caption{}
        \includegraphics[width=\linewidth]{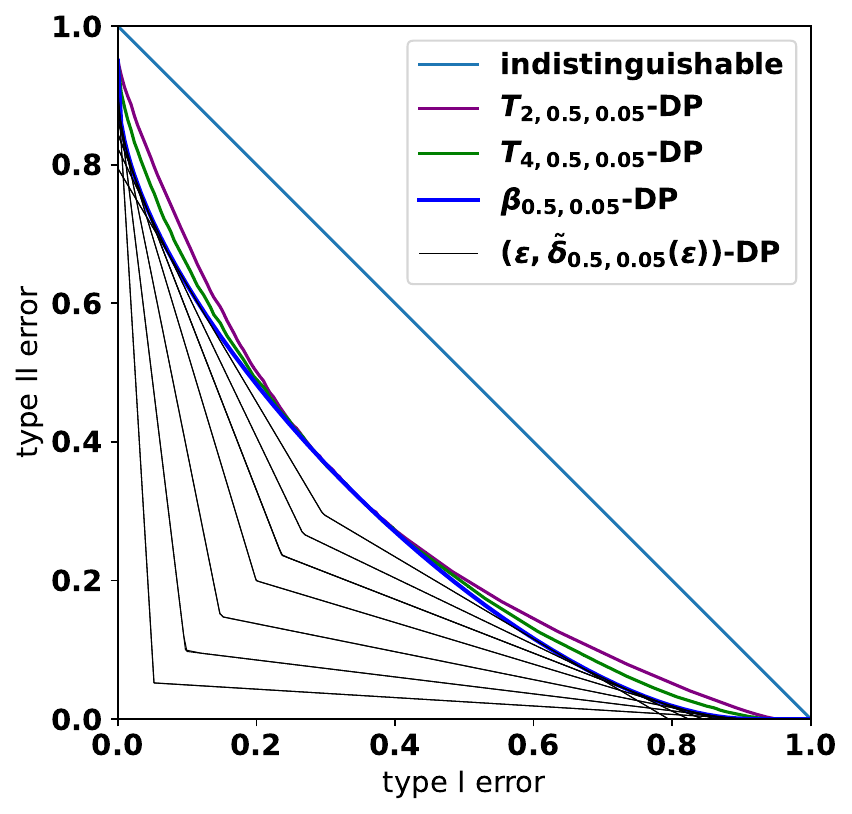}
        \label{fig:envelopes}
    \end{subfigure}
    \quad
    \begin{subfigure}[b]{0.45\textwidth}
        \caption{}
        \includegraphics[width=\linewidth]{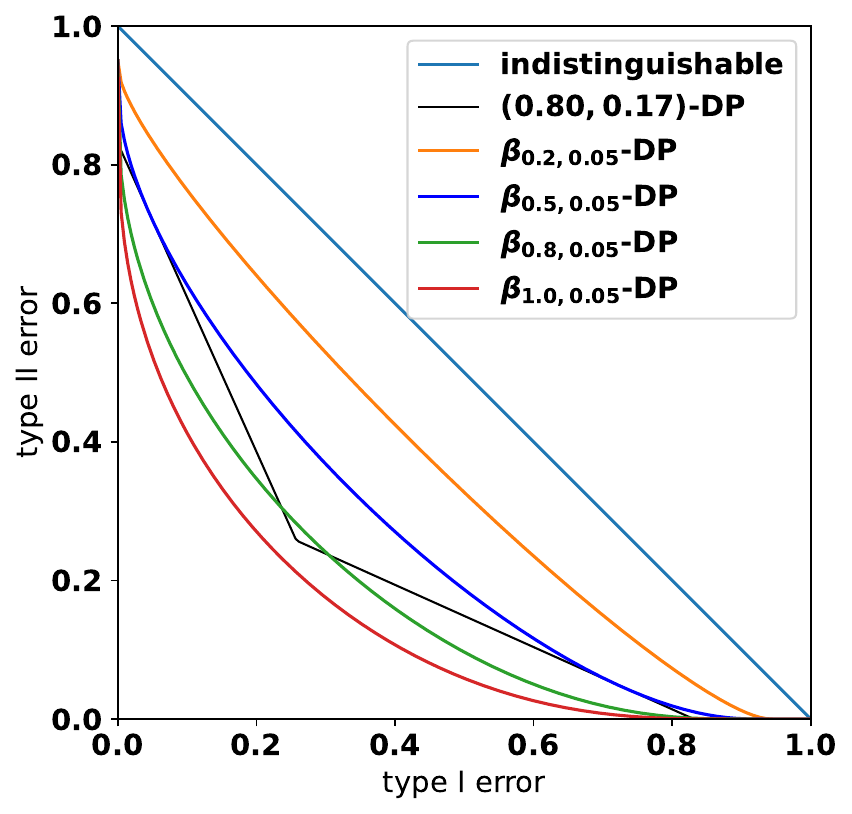}
        \label{fig:find_c}
    \end{subfigure}
    \caption{Asymptotic trade-off function $\beta_{0.5,0.05}(\alpha)$ (blue) along with a set of its $(\varepsilon, \tilde{\delta}_{0.5,0.05}(\varepsilon))$-DP envelop (black polylines) for $\varepsilon=0.5,\ 0.7,\ 0.9,\ 1.2,\ 1.6,\ 2.1,\ 2.8$ and the trade-off functions $T_{2,0.5,0.05}$ (purple) and $T_{4,0.5,0.05}$ (green) evaluated by $10^{5}$ simulations in Panel (a). 
    The $(0.8,0.17)$-DP polyline (black) along with four $\beta_{c,0.05}(\alpha)$ curves for $c=0.2\text{ (orange)},0.5\text{ (blue)},0.8\text{ (green)}$ and $1\text{ (red)}$ in Panel (b). 
   The indistinguishable curve $\{(\alpha,1-\alpha):\alpha\in[0,1]\}$ (cyan)  is also marked.}  \label{fig:1}
\end{figure}

\Cref{thm:profile} offers a practical guidance for  finding \((\delta, \lambda)\) such that \(\mathrm{Mech}_{\text{ZIL}}(\cdot; \delta, \lambda)\) achieves a specified \((\varepsilon', \delta')\)-ADP or \((\varepsilon', \delta')\)-DP. First of all, for given $\varepsilon'$, $\delta'$ and $\delta$, we solve \(c'\) such that \(\tilde{\delta}_{c', \delta}(\varepsilon') = \delta'\). Then, $f_{\varepsilon', \delta'}$ is dominated by \(\beta_{c', \delta}\). 
Secondly, choose $\lambda = \max_{j \in \{1, \ldots, d\}} \mathrm{diam}(\mathcal{X}_{j}) / c'$ for \((\varepsilon', \delta')\)-ADP and $\lambda = \mathrm{diam}(\mathcal{X}) / c'$ for \((\varepsilon', \delta')\)-DP. The resulting ZIL mechanism can guarantee the required privacy level.
Note that the above choice of $\lambda$ depends on the choice of $\delta$, and the solution $c'$ exists if and only if $\delta<\delta'$. Furthermore, since \(\beta_{c}(\alpha)\) is decreasing as $c$ increases, 
for any \(c \leqslant c'\), we have \(\beta_{c, \delta}(\alpha) \geqslant \beta_{c', \delta}(\alpha) \geqslant f_{\varepsilon', \delta'}(\alpha)\). 
\Cref{fig:1}(b) demonstrate the above procedure for $\varepsilon' = 0.8$, $\delta' = 0.17$, and $\delta = 0.05$, where $c'=0.5$ solves $\tilde{\delta}_{c',\delta}(\varepsilon')=\delta'$. It is seen from \Cref{fig:1}(b) that $\beta_{0.5,0.05}$ precisely takes $f_{0.8,0.17}$ as its envelope.  
}

Although we have only proven that the ZIL mechanism satisfies the central differential privacy in the presence of a curator, who holds all the data,  the ZIL mechanism 
also satisfies the local differential privacy (LDP) \citep{kasiviswanathan2011can} at the same level. This is because the noise-adding step can be completed on the individual's side, without having to wait until the data are loaded to the curator.

\section{Versatile Differentially Private M-Estimation} 
\label{sec:M-estimation}

{In this section, we introduce the estimation procedure based on the noisy data from the proposed differentially private ZIL mechanism, demonstrate its versatility for a general class of M-estimation that does not require the smoothness of the loss function, and explain its advantages for statistical inference.} 

\subsection{Doubly Random Corrected Loss}
We consider the M-estimation in a semiparametric framework for general statistical inference. 
Let $\ell(x,\theta)$ be a loss function specified by an analyst with $x\in\mathbb{R}^{d}$ and $\theta\in\Theta\subset\mathbb{R}^{p}$. Suppose that the original data $\{X_{i}\}_{i=1}^{n}$ are drawn from a distribution $F$ with a compact support $\mathcal{X}\subset\mathbb{R}^{d}$.
Let $\tilde{X}_{i}^{(1)} = X_{i} + Z_{i}$ be the noisy data from the ZIL mechanism in \Cref{def:ZIL}, where $Z_{1}, \ldots, Z_{n}$ are i.i.d. from the distribution $\text{ZIL}(\delta,\lambda^{2}I_{d})$.
The task of the analyst is to estimate the true parameter $\theta_{0} = \operatorname{argmin}_{\theta\in\Theta} \E \ell(X_{i},\theta)$ based on the noisy data $\{\tilde{X}_{i}^{(1)}\}_{i = 1}^{n}$. The ZIL mechanism allows the analyst to choose any form of the loss function of interest under minimum regularity conditions, as demonstrated in the following.

Estimation of $\theta$ is related to parameter estimation under data with measurement error. If the original data $\{X_{i}\}_{i=1}^{n}$ were observable, one could attain the oracle M-estimator 
\begin{equation}\label{eq:oracle}
\hat{\theta}_{n}^{\text{ORA}} = \underset{\theta\in\Theta}{\operatorname{argmin}} \sum_{i=1}^{n}\ell(X_{i},\theta).
\end{equation} 
Replacing the original data with the noisy data 
 $\{\tilde{X}_{i}^{(1)}\}_{i=1}^{n}$ 
in \eqref{eq:oracle} results in the naive estimator 
\begin{equation}\label{eq:naive}
    \hat{\theta}_{n}^{\text{NAI}} = \underset{\theta\in\Theta}{\operatorname{argmin}} \sum_{i=1}^{n} \ell(\tilde{X}_{i}^{(1)}, \theta), 
\end{equation}
which may be inconsistent to $\theta_{0}$ as there is no guarantee that $n^{-1}\sum_{i=1}^{n}\ell(\tilde{X}_{i}^{(1)},\theta)$ is unbiased to the underlying risk  $\E \ell(X_{i},\theta)$.
To obtain a consistent estimator, a corrected loss function needs to be constructed which requires $\ell(x, \theta)$ to be sufficiently smooth with respect to $x$ or necessitates truncated Fourier transform by numerical integration. 

The use of the ZIL noise can avoid these issues and bring a new method for consistent parameter estimation. 
We propose a double random DP mechanism (DRDP)
that adds  additional SL noises on the released data $\{\tilde{X}_{i}^{(1)}\}_{i = 1}^{n}$. Let 
$$\tilde{X}_{i}^{(2)}=\tilde{X}_{i}^{(1)}+S_{i}$$ for $i=1,\ldots,n$, where $\{S_{i}\}_{i=1}^{n}$ are i.i.d. from the distribution $\mathcal{SL}_{d}(\delta\lambda^{2}I_{d})$, as shown in Algorithm 1.
Using the two sets of privacy-protected data \(\{\tilde{X}_{i}^{(1)}\}_{i=1}^{n}\) and \(\{\tilde{X}_{i}^{(2)}\}_{i=1}^{n}\), we define a doubly random corrected loss (DRCL)
\begin{equation}\label{def:DRCL} 
\ell^{\text{DR}}(\tilde{X}_{i}^{(1)}, \tilde{X}_{i}^{(2)}, \theta;\delta) = \left(1 - \delta^{-1}\right) \ell(\tilde{X}_{i}^{(2)}, \theta) + \delta^{-1} \ell(\tilde{X}_{i}^{(1)}, \theta). 
\end{equation}
The term ``doubly random'' (DR) comes from the use of both $\{\tilde{X}_{i}^{(1)}\}$ and $\{\tilde{X}_{i}^{(2)}\}$
to implicitly correct the loss function. 
Indeed, \Cref{thm:identification} shows that $\E \ell(X_{i}, \theta) = \E \ell^{\text{DR}}(\tilde{X}_{i}^{(1)}, \tilde{X}_{i}^{(2)}, \theta; \delta)$, indicating the DRCL is unbiased to the loss function $\ell(X_{i}, \theta)$ with the original data.

The differentially private DRCL estimator is defined as 
\begin{equation}\label{def:DRCL-estimator}
\hat{\theta}_{n}^{\text{DR}} = \underset{\theta\in\Theta}{\operatorname{argmin}}\sum_{i=1}^{n}\ell^{\text{DR}}(\tilde{X}_{i}^{(1)}, \tilde{X}_{i}^{(2)}, \theta;\delta). 
\end{equation}
The advantage of the DRCL is that it does not involve any differentiation of the loss $\ell(x, \theta)$ or numerical integration, which lowers the requirement on the loss function and reduces computation complexity.
This is especially useful for complex or non-smooth loss functions, such as those in the neural networks where ReLU active functions are involved. In Section \ref{sec:simu-non-smooth}, we show that the existing methods for correcting measurement error bias cannot be applied to the ReLU function, whereas the proposed DRCL method remains valid. 

{In the following, we provide an explanation for the rationale of constructing the DRCL under the special case that $\ell(x, \theta)$ is twice-differentiable with respect to $x$. As $S_{i} \sim \mathcal{SL}_d(\delta \lambda^2 I_d)$, \Cref{lemma:CL_for_SL} in the SM shows that 
\be\label{eq:explain-2}
\E\ell(\tilde{X}_{i}^{(1)},\theta) = \E\ell(\tilde{X}_{i}^{(2)},\theta)-\frac{\delta\lambda^{2}}{2}\sum_{k=1}^{d} \E \frac{\partial^{2}}{\partial x_{k}^{2}}\ell(\tilde{X}_{i}^{(2)},\theta).
\ee
As $Z_{i}\sim \ZILdis(\delta,\lambda^{2}I_{d})$ and $Z_{i}+S_{i} \sim \mathcal{SL}_d(\lambda^2 I_d)$, \Cref{lemma:CL_for_SL} again implies that 
\be\label{eq:explain-1}
\E\ell(X_{i},\theta) = \E\ell(\tilde{X}_{i}^{(2)},\theta)-\frac{\lambda^{2}}{2}\sum_{k=1}^{d} \E \frac{\partial^{2}}{\partial x_{k}^{2}}\ell(\tilde{X}_{i}^{(2)},\theta).
\ee
Combining \eqref{eq:explain-2} and \eqref{eq:explain-1}, it leads to $\E \ell^{\text{DR}}(\tilde{X}_{i}^{(1)}, \tilde{X}_{i}^{(2)}, \theta;\delta) = \E \ell(X_{i},\theta)$, which shows the unbiasedness of the DRCL. Although the above derivation is based on the existence of $\partial^2 \ell(x, \theta) / \partial x_k^2$ for all $k = 1, \ldots, d$, $\ell^{\text{DR}}(\tilde{X}_{i}^{(1)}, \tilde{X}_{i}^{(2)}, \theta;\delta)$ does not involve any differentiation, and the same conclusion even holds for 
$\ell(x, \theta)$ with some discontinuity as shown in \Cref{thm:identification}. 
The following algorithm shows the procedure of the DRDP mechanism that facilitates the DRCL estimation.

\begin{center}
\begin{algorithm}[H]\label{alg:data_pre}
\caption{Doubly random differentially private (DRDP) mechanism}
\KwIn{original dataset $\{X_{i}\}_{i=1}^n$, privacy parameters $\delta$ and $\lambda$ of the ZIL mechanism. 
}
        \textbf{Step 1}:      
        generate the ZIL noisy data \(\{\tilde{X}_{i}^{(1)}\}_{i=1}^{n}=\text{Mech}_{\text{ZIL}}(\{X_{i}\}_{i=1}^n;\delta,\lambda)\);\

        \textbf{Step 2}:         
        generate  the 
        doubly randomized data $\tilde{X}_{i}^{(2)}=\tilde{X}_{i}^{(1)}+S_{i}$ for $i=1,\ldots,n$ where $\{S_{i}\}_{i=1}^{n}$ are i.i.d. $\mathcal{SL}_{d}(\delta\lambda^{2}I_{d})$. 
        
	\KwOut{$\{\tilde{X}_{i}^{(1)}\}_{i=1}^{n}$, $\{\tilde{X}_{i}^{(2)}\}_{i=1}^{n}$ and $\delta$.}
\end{algorithm}
\end{center}

Algorithm \ref{alg:data_pre} is a differentially private mechanism. It takes a dataset $\mathbf{X} = \{X_i\}_{i=1}^{n}$ as input and returns the output of the ZIL mechanism $\{\tilde{X}_{i}^{(1)}\}_{i=1}^{n}$, along with its privacy parameter $\delta$ and a post-processed product, the doubly randomized dataset $\{\tilde{X}_{i}^{(2)}\}_{i=1}^{n}$. 
Note that the DRCL for any loss function can be computed using the outputs of the DRDP mechanism.  
According to the post-processing property in \Cref{thm:post-processing}, the privacy protection capability of the DRDP mechanism in Algorithm \ref{alg:data_pre} is fully inherited from the privacy protection capability of the ZIL mechanism in \Cref{def:ZIL}.

The following proposition provides the privacy guarantees of the DRDP mechanism. 

\begin{proposition}\label{prop:dp_guarantee_alg}
     Suppose that $\{X_i\}_{i=1}^{n}$ are real-valued vectors from a $d$-dimensional distribution with a compact support $\mathcal{X}$. 
     {The DRDP mechanism %
     satisfies \(T_{d,c_{A},\delta}\)-ADP and \(T_{d,c_{I},\delta}\)-DP for $c_{A} = \max_{j\in\{1,\ldots,d\}}\mathrm{diam}(\mathcal{X}_{j})/\lambda$ and $c_{I} = \mathrm{diam}(\mathcal{X})/\lambda$, respectively.}
\end{proposition}

\subsection{Connection to Measurement Error Literature}\label{sec:connection}

Parameter estimation using privacy-protected data with added noises is well connected to the measurement error problem. The deconvolution method is an important method in the measurement error literature. 
\cite{stefanski_deconvolving_1990} constructed kernel deconvolution estimators for the underlying density function. \cite{wang_correctedloss_2012} derived a corrected loss by first smoothing the check loss and then applying the deconvolution and Fourier inversion in the context of quantile regression with noisy covariates. \cite{firpo_measurement_2017} also considered the scenario of quantile regression with noisy covariates, where they first use the deconvolution approach to estimate the density function of the authentic data, and then substitute it into the estimating equation to solve for the parameters. \cite{ran_yang_density_2020} provided a density estimator for noisy data by solving a linear system corresponding to the deconvolution problem. \cite{kent_smoothness-penalized_2023} provided the convergence rate of the estimator in \cite{ran_yang_density_2020}.

We now present a formulation of the deconvolution approach in the general context of the M-estimation that is designed to estimate the true parameter $\theta_{0} = \operatorname{argmin}_{\theta\in\Theta} \E \ell(X_{i},\theta)$ based on the noisy data $\{\tilde{X}_{i}\}_{i = 1}^{n}$, where $\tilde{X}_{i} = X_i + \tilde{Z}_{i}$ and $\{\tilde{Z}_{i}\}_{i = 1}^{n}$ are i.i.d. noises. Let \(\varphi_{\ell,\theta}(t) = \int \ell(x,\theta) e^{it^{\top}x} \rd x\) be the Fourier transform of the loss \(\ell(x,\theta)\), and \(\varphi_{\tilde{z}}(t)\) be the characteristic function of \(\tilde{Z}_{i}\). 
Suppose that \(\ell(x, \theta)\) is continuous and integrable with respect to \(x\), \(\varphi_{\ell, \theta}(t)\) is integrable, and \(\mathbb{E}|\ell(X, \theta)| < \infty\). 
Then, by applying Fourier inversion and Fubini theorem, the following result can be obtained:
\begin{equation}\label{eq:general-result}
     \mathbb{E} \ell(X,\theta) = \int\int \frac{1}{(2\pi)^{d}} e^{-it^{\top}x} \frac{\varphi_{\ell,\theta}(t)}{\varphi_{\tilde{z}}(-t)} \, \rd t \, \rd F_{\tilde{X}}(x).
\end{equation}
which %
implies a corrected loss function 
\begin{equation}\label{eq:Corrected Loss}
 \tilde{\ell}(x,\theta) = \frac{1}{(2\pi)^{d}}\int e^{-it^{\top}x}\frac{\varphi_{\ell,\theta}(t)}{\varphi_{\tilde{z}}(-t)}\rd t.
\end{equation}
If the integral exists, 
indicating the underlying risk can be recovered by the corrected risk. However, the existence of the integral $\int e^{-it^{\top}x} \varphi_{\ell,\theta}(t) / \varphi_{\tilde{z}}(-t) \rd t$ requires restrictive conditions, which may not be satisfied for many loss functions and error distributions, for example, the $L_2$ loss with the Gaussian error. A remedy for this problem is to 
truncate the integral in \eqref{eq:Corrected Loss} or multiply a rapidly decaying characteristic function in the numerator to counteract the divergence of $1/\varphi_{\tilde{z}}(t)$ as $t \to \infty$.

In comparison, the proposed DRCL only requires mild conditions on the loss function $\ell(x, \theta)$ and is free from any condition on the Fourier transform $\varphi_{\ell,\theta}(t)$ in the frequency domain. Additionally, it does not require numerical integration or selection of hyper-parameters for evaluating the corrected loss in \eqref{eq:Corrected Loss} as in the case of the deconvoluted kernel density estimation where a smoothing parameter is needed.
These make the proposed DRCL more generally applicable and computationally more efficient. 
Another approach based on \eqref{eq:Corrected Loss} 
to construct the corrected loss function is based on differentiation of the loss $\ell(x, \theta)$, which requires sufficient smoothness of $\ell(x, \theta)$ with respect to $x$. 
Suppose the components of $\tilde{Z}_{i}$ are independent Laplace random variables, where $\tilde{Z}_{i} = (\tilde{Z}_{i1}, \ldots, \tilde{Z}_{id})^{\top}$ and $\tilde{Z}_{ij}$ follows the Laplace distribution with variance $\lambda_{j}^2$ for $j = 1, \ldots, d$. Applying the derivative theorem in Fourier transformation to (\ref{eq:Corrected Loss}), the corrected loss under component-wise independent Laplace noise is 
\begin{equation}\label{eq:CL-Lap}
\tilde{\ell}^{\text{L}}(x,\theta)=\prs{1-\frac{\lambda_{1}^{2}}{2}\frac{\partial^{2}}{\partial x_{1}^{2}}}\ldots \prs{1-\frac{\lambda_{d}^{2}}{2}\frac{\partial^{2}}{\partial x_{d}^{2}}} \ell(x,\theta).
\end{equation}
If $\tilde{Z}_{i}$ follows the SL distribution $\mathcal{SL}_{d}(\lambda^{2}I_{d})$, \Cref{lemma:CL_for_SL} in the SM provides the corresponding corrected loss
\begin{equation}\label{eq:CL_SL}
\tilde{\ell}^{\text{SL}}(x,\theta)=\ell(x,\theta)-\frac{\lambda^{2}}{2}\sum_{k=1}^{d}\frac{\partial^{2}}{\partial x_{k}^{2}}\ell(x,\theta),
\end{equation} 
which only depends on the second-order derivatives. 
Although the above forms of corrected functions avoid numerical integration, they require the loss function to be sufficiently smooth with respect to $x$, which is not satisfied by the quantile regression or the ReLU activation function. Whereas, the proposed method does not have such restrictions. 

\cite{novick2002corrected} considered estimating a nonlinear function of mean under Gaussian measurement error by 
further adding complex-valued Gaussian random variables to the noisy data. 
However, they require their nonlinear function to be 
differentiable everywhere in the complex plane,
which is a more restrictive requirement. The loss function of the logistic regression does not satisfy this condition because $1/(1+e^{-x})$ 
has singular points in the complex plane. The proposed DRCL does not require such conditions and operates for the logistic regression.

In summary, the proposed DRCL is more generally applicable 
than the existing methods that handle measurement errors. 
It works for a general class of loss functions under minimum conditions. 
A key innovation of the DRDP mechanism is introducing additional SL noises on the ZIL noises added for privacy protection (Step 1 of Algorithm \ref{alg:data_pre}). Both noises are designed to replace the differentiation and numerical integration in the existing corrected loss formulation.

\section{Properties of Doubly Random  Corrected M-Estimation} 
\label{sec:DRCL}
The study on the consistency and the asymptotic normality of the DRCL estimator requires the following conditions. Let $\mu^{(1)}$ and $\mu^{(2)}$ be two induced measures on $\mathbb{R}^{d}$ by $\tilde{X}_i^{(1)}$ and $\tilde{X}_i^{(2)}$, respectively.

\begin{condition}\label{cond:identification}
    (i) For $\theta\in\Theta$, $\E|\ell(X_{i},\theta)|<\infty$, $\E|\ell(\tilde{X}_{i}^{(1)},\theta)|<\infty$ and $\E|\ell(\tilde{X}_{i}^{(2)},\theta)|<\infty$. (ii) For $\theta\in\Theta$, $\ell(x,\theta)$ viewed as a function of $x$ has a set of discontinuities denoted by $D_{\theta}$. Assume that for every $\theta\in\Theta$, $\Prob(X\in D_{\theta})=0$, and for any bounded set $B\in\mathbb{R}^{d}$, the intersection $D_{\theta}\cap B$ is finite.
\end{condition}

\begin{condition}
    \label{cond:consistency}
    (i) The parameter space $\Theta$ is a compact set, and the true parameter $\theta_{0}$ is an interior point of $\Theta$. For any $x$, $\ell(x, \theta)$ is continuous with respect to $\theta$.\label{cond: space}
        (ii) Uniform law of large numbers: $\sup_{\theta\in\Theta}| n^{-1} \sum_{i=1}^{n}\ell(\tilde{X}_{i}^{(2)},\theta)-\E \ell(\tilde{X}_{i}^{(2)},\theta)|\overset{p}{\to}0$ and $\sup_{\theta\in\Theta}| n^{-1} \sum_{i=1}^{n}\ell(\tilde{X}_{i}^{(1)},\theta)-\E \ell(\tilde{X}_{i}^{(1)},\theta)|\overset{p}{\to}0$ as $n\to\infty$.
        (iii) Separability: for every $\varepsilon>0$, $\inf_{\theta:d(\theta,\theta_{0})\geqslant\varepsilon}\E \ell(X_{i},\theta)>\E \ell(X_{i},\theta_{0})$. \label{cond: sepa}
\end{condition}

\begin{condition}\label{cond:AN}
    (i) Assume $\ell(x,\theta)$ is differentiable at $\theta_{0}$ with derivative $\nabla_{\theta} \ell(x,\theta_{0})$ almost surely for $\mu^{(1)}$ and $\mu^{(2)}$, and
    \begin{equation}
        |\ell(x,\theta_{1})-\ell(x,\theta_{2})|\leq \dot{\ell}(x)\|\theta_{1}-\theta_{2}\|
    \end{equation}    
    for every $\theta_{1}$ and $\theta_{2}$ in a neighborhood of $\theta_{0}$ and a measurable function $\dot{\ell}$ with $\E\dot{\ell}(\tilde{X}_{i}^{(2)})^{2}<\infty$ and $\E\dot{\ell}(\tilde{X}_{i}^{(1)})^{2}<\infty$.
    (ii) Assume that the map $\theta\mapsto\E \ell(X_{i},\theta)$ admits a second-order Taylor expansion at $\theta_{0}$ with nonsingular symmetric second derivative matrix $V_{\theta_{0}}$.  
\end{condition}

Condition \ref{cond:identification} is used to establish the unbiasedness of \( \ell^{\text{DR}}(\tilde{X}_{i}^{(1)}, \tilde{X}_{i}^{(2)}, \theta) \). The integrability assumption in Condition \ref{cond:identification} (i) is fairly mild. Since both the ZIL and SL noises have finite second moments and the original observation $X_i$ is bounded, this assumption is satisfied for the $L_{2}$ and check losses.
Condition \ref{cond:identification} (ii) allows dis-continuity points of $\ell(x,\theta)$ with respect to $x$, which is much weaker than the sufficient smoothness condition of $\ell(x, \theta)$ with respect to $x$ required by the deconvolution approach for loss adjustment. 
The loss functions in the contexts of linear regression, logistic regression, quantile regression, and neural networks all satisfy Condition \ref{cond:identification} (ii). To establish the asymptotic normality of the DRCL estimator, we require Conditions \ref{cond:consistency} and \ref{cond:AN}, which are standard assumptions for the M-estimation, as outlined in \citet{vaart_asymptotic_2007}. For example, the $\tau$-quantile estimation task $\theta_{0} = \arg\min_{\theta \in \Theta} \E (X_{i} - \theta)(\tau - \mathbf{1}(X_{i} - \theta < 0))$, along with the output of Algorithm \ref{alg:data_pre}, satisfies those three conditions provided that $\Theta$ is tight, $\E|X_{i}|<\infty$, $F'(\theta_{0}) > 0$ and $F''(\theta)$ exists in a neighborhood of $\theta_{0}$ with $F''(\theta_{0}) \neq 0$. The example involving the ReLU function $\theta_{0} = \arg\min_{\theta \in \Theta} \E (\theta - \text{ReLU}(X_{i}))^{2}$ considered in the simulation study also satisfies the three conditions provided that $\Theta$ is tight and $\E X_{i}^{2}<\infty$.

\begin{theorem}\label{thm:identification}
Under Condition \ref{cond:identification}, with the noisy data $\{\tilde{X}_{i}^{(1)}\}_{i=1}^{n}$ and $\{\tilde{X}_{i}^{(2)}\}_{i=1}^{n}$ obtained in Algorithm \ref{alg:data_pre}, and the privacy parameter $\delta$,
it follows that for every \(\theta \in \Theta\), 
\begin{equation}\label{eq:identification}
\E \ell(X_{i}, \theta) = \E \ell^{\text{DR}}(\tilde{X}_{i}^{(1)}, \tilde{X}_{i}^{(2)}, \theta; \delta).
\end{equation}
\end{theorem}

\Cref{thm:identification} suggests that  $\ell^{\text{DR}}$ acts as a surrogate for the underlying loss $\ell$, which makes the differentially private DRCL estimator \eqref{def:DRCL-estimator} consistent to the underlying $\theta$. Compared to the existing approaches to obtain a corrected loss in Section \ref{sec:connection}, which require the existence of higher-order derivatives of \(\ell(x, \theta)\) with respect to \(x\) or necessitate numerical integration, the DRCL with the DRDP noise only requires that the loss \(\ell(x, \theta)\) is continuous with respect to \(x\). In addition, the DRCL only uses basic arithmetic operations 
when calculating the loss \(\ell\), making it computationally very efficient.

The DRCL estimator is differentially private, as it is based on the output of the DRDP mechanism in Algorithm \ref{alg:data_pre}. According to \Cref{prop:dp_guarantee_alg} and the post-processing property (\Cref{thm:post-processing}), $\hat{\theta}_{n}^{\text{DR}}$ satisfies $T_{d,c_{A},\delta}$-ADP and $T_{d,c_{I},\delta}$-DP for $c_{A} = \max_{1 \leq j \leq d}\mathrm{diam}(\mathcal{X}_{j})/\lambda$ and $c_{I} = \mathrm{diam}(\mathcal{X})/\lambda$. 

The following theorems present the consistency and the asymptotic normality of $\hat{\theta}_{n}^{\text{DR}}$. 

\begin{theorem}\label{thm:consistency}
    Under Conditions \ref{cond:identification} and \ref{cond:consistency}, 
    $\hat{\theta}_{n}^{\text{DR}}\overset{p}{\to}\theta_{0}$ as $n\to\infty$.
\end{theorem}

\begin{theorem}\label{thm:normality}
    Under Conditions \ref{cond:identification}, \ref{cond:consistency} and \ref{cond:AN}, 
        $\sqrt{n}(\hat{\theta}_{n}^{\text{DR}}-\theta_{0})\overset{d}{\to}N(0,V_{\theta_{0}}^{-1}A(\delta,\lambda)V_{\theta_{0}}^{-1})$,
   as $n\to \infty$ where 
   $A(\delta,\lambda)=\E\big\{\nabla_{\theta}\ell^{\text{DR}}(\tilde{X}_{i}^{(1)}, \tilde{X}_{i}^{(2)}, \theta_{0};\delta)\} \big\{\nabla_{\theta}\ell^{\text{DR}}(\tilde{X}_{i}^{(1)}, \tilde{X}_{i}^{(2)}, \theta_{0};\delta)\}^{\top}$.
\end{theorem}

The M-estimator on the original data (the oracle estimator $\hat{\theta}_{n}^{\text{ORA}}$) has the asymptotic variance 
\[
V_{\theta_{0}}^{-1}\E \nabla_{\theta}\ell(X_{i},\theta_{0})\nabla_{\theta}\ell(X_{i},\theta_{0})^{\top}V_{\theta_{0}}^{-1}.
\]
The difference between the asymptotic variances of the Oracle estimator and the DRCL estimator lies in the middle term of the sandwich form.
In the following corollary, we prove that under mild conditions, 
\[
A(\delta,\lambda)\geqslant \E \nabla_{\theta}\ell(X_{i},\theta_{0})\nabla_{\theta}\ell(X_{i},\theta_{0})^{\top},
\]
indicating a cost in the estimation efficiency due to privacy protection.  
Moreover, the equality holds whenever \(\delta = 1\) or \(\lambda = 0\). In both cases, the two noisy datasets, \(\{\tilde{X}_{i}^{(1)}\}_{i=1}^{n}\) and \(\{\tilde{X}_{i}^{(2)}\}_{i=1}^{n}\), degenerate to the true dataset \(\{X_{i}\}_{i=1}^{n}\). %

\begin{corollary}\label{thm:efficiency} 
    Suppose that \(\nabla_{\theta} \ell(x, \theta_{0})\), viewed as a vector-valued function of \(x\), has a set of discontinuities denoted by \(D_{\theta_{0}}\). Assume that \(\Prob(X \in D_{\theta_{0}}) = 0\), and for any bounded set \(B \subset \mathbb{R}^{d}\), the intersection \(D_{\theta_{0}} \cap B\) is finite.
 Then, $A(\delta,\lambda)\geqslant \E \nabla_{\theta}\ell(X_{i},\theta_{0})\nabla_{\theta}\ell(X_{i},\theta_{0})^{\top}$.
\end{corollary} 

\cite{avella-medina_differentially_2023} showed that the M-estimator derived from Noisy-SGD \citep{rajkumar_differentially_2012} 
has the same asymptotic variance as the oracle estimator. 
However, compared to the DRCL estimator, the Noisy-SGD is not versatile, since it is designed for a pre-specified task and can only be employed once on a dataset, and cannot be adapted to newly arrived data and tasks. The definition of versatile differential privacy mechanism is in the introduction.
The differentially private M-estimator based on the Perturbed-Histogram \citep{lei_differentially_2011}, in general, has a slower convergence rate than $1/\sqrt{n}$ and lacks asymptotic normality results, while requiring a smoothing parameter. More discussions about Noisy-SGD and Perturbed-Histogram are provided in Section \ref{sec:discuss}.

\section{Smoothed Double Random Estimator}  %
\label{sec:alternatives} 

In addition to the DRCL formulation which does not require the loss function to be smooth,  we propose a differentially private M-estimation 
if the loss is known to be smooth up to the second order. 

Since $\{S_{i}\}_{i=1}^{n}$ and $\{Z_{i}+S_{i}\}_{i=1}^{n}$ are both SL distributed but with different variances, applying \eqref{eq:CL_SL} twice and recalling that $\tilde{X}_{i}^{(1)} = \tilde{X}_{i} + Z_i$ and $\tilde{X}_{i}^{(2)} = \tilde{X}_{i}^{(1)} + S_i$, we have  
\begin{align}
    \E \ell(X_{i},\theta) &= \E \prs{\ell(\tilde{X}_{i}^{(2)},\theta) - \frac{\lambda^{2}}{2}\sum_{k=1}^{d}\frac{\partial^{2}}{\partial x_{k}^{2}}\ell(\tilde{X}_{i}^{(2)},\theta)} \label{eq:first-eq} \quad \hbox{and}\\
    \E \ell(\tilde{X}_{i}^{(1)},\theta) &= \E\prs{\ell(\tilde{X}_{i}^{(2)},\theta)-\frac{\delta\lambda^{2}}{2}\sum_{k=1}^{d}\frac{\partial^{2}}{\partial x_{k}^{2}}\ell(\tilde{X}_{i}^{(2)},\theta)},\label{eq:second-eq}
\end{align}
if $\ell(x, \theta)$ is twice differentiable with respect to $x$.
Substitute \eqref{eq:second-eq} to \eqref{eq:first-eq}, we obtain an unbiased recovery of the underlying expected loss: 
\begin{equation}
\ell^{\text{SDR}}(\tilde{X}_{i}^{(1)},\tilde{X}_{i}^{(2)},\theta;\delta,\lambda) = \ell(\tilde{X}_{i}^{(1)},\theta)-\frac{(1-\delta)\lambda^{2}}{2}\sum_{k=1}^{d}\frac{\partial^{2}}{\partial x_{k}^{2}}\ell(\tilde{X}_{i}^{(2)},\theta).
\end{equation}
This leads to a smoothed doubly random corrected loss (sDRCL) estimator 
\begin{equation}\label{eq:def-sDRCL}
\hat{\theta}_{n}^{\text{SDR}} = \underset{\theta\in\Theta}{\operatorname{argmin}} \sum_{i=1}^{n}\prb{\ell(\tilde{X}_{i}^{(1)},\theta)-\frac{(1-\delta)\lambda^{2}}{2}\sum_{k=1}^{d}\frac{\partial^{2}}{\partial x_{k}^{2}}\ell(\tilde{X}_{i}^{(2)},\theta)}.
\end{equation}
The sDRCL may be seen as a version of the DRCL for situations where the loss is smooth to the second order. 

From \eqref{eq:first-eq}, %
one may have another unbiased corrected loss %
\begin{equation}\label{eq:CL}
\ell^{\text{SL}}(\tilde{X}_{i}^{(2)},\theta;\lambda) = \ell(\tilde{X}_{i}^{(2)},\theta)-\frac{\lambda^{2}}{2}\sum_{k=1}^{d}\frac{\partial^{2}}{\partial x_{k}^{2}}\ell(\tilde{X}_{i}^{(2)},\theta) 
\end{equation} 
by using only one data set $\{\tilde{X}_{i}^{(2)}\}_{i = 1}^{n}$,
which leads to the SL corrected-loss estimator
\begin{equation}\label{eq:def-cl} 
\hat{\theta}^{\text{SL}}_{n} = \underset{\theta\in\Theta}{\operatorname{argmin}} \sum_{i=1}^{n}\prb{\ell(\tilde{X}_{i}^{(2)},\theta)-\frac{\lambda^{2}}{2}\sum_{k=1}^{d}\frac{\partial^{2}}{\partial x_{k}^{2}}\ell(\tilde{X}_{i}^{(2)},\theta)}.
\end{equation} 

The following theorem  establish the asymptotic normality of $\hat{\theta}^{\text{SL}}$ and $\hat{\theta}^{\text{SDR}}$, respectively. 
\begin{theorem}\label{thm:AN-CL}
    Suppose \Cref{cond:consistency} (i), \ref{cond:consistency} (iii) and \ref{cond:AN} (ii) hold. (i) If  \Cref{cond:AN-sDRCL} in the SM holds,  
        $\sqrt{n}(\hat{\theta}_{n}^{\text{SDR}}-\theta_{0})\overset{d}{\to}N(0,V_{\theta_{0}}^{-1}A^{\text{SDR}}(\delta,\lambda)V_{\theta_{0}}^{-1})$,
    where $$A^{\text{SDR}}(\delta,\lambda)=\E \big\{\nabla_{\theta}\ell^{\text{SDR}}(\tilde{X}_{i}^{(1)},\tilde{X}_{i}^{(2)},\theta_{0};\delta,\lambda)\big\} \big\{\nabla_{\theta}\ell^{\text{SDR}}(\tilde{X}_{i}^{(1)},\tilde{X}_{i}^{(2)},\theta_{0};\delta,\lambda)\big\}^{\top}.$$ 
    
    (ii) If \Cref{cond:AN-CL} in the SM holds,  
        $\sqrt{n}(\hat{\theta}_{n}^{\text{SL}}-\theta_{0})\overset{d}{\to}N(0,V_{\theta_{0}}^{-1}A^{\text{SL}}(\lambda)V_{\theta_{0}}^{-1})$,
    where $$A^{\text{SL}}(\lambda)=\E \big\{\nabla_{\theta}\ell^{\text{SL}}(\tilde{X}_{i}^{(2)},\theta_{0};\lambda)\big\} \big\{\nabla_{\theta}\ell^{\text{SL}}(\tilde{X}_{i}^{(2)},\theta_{0};\lambda)\big\}^{\top}.$$    
\end{theorem}

Similar to \Cref{cond:identification}--\Cref{cond:AN}, \Cref{cond:AN-sDRCL} and \Cref{cond:AN-CL}  are the standard regularity conditions needed for the asymptotic normality of the SL and sDRCL estimators, respectively. They include the continuity condition of the second-order derivative of $\ell(x, \theta)$ with respect to $x$ for the the unbiasedness of $\ell^{\text{SDR}}$ and $\ell^{\text{SL}}$ and the standard M-estimation conditions applied to $\ell^{\text{SDR}}$ and $\ell^{\text{SL}}$.

There is no clear ordering in general regarding the asymptotic variances among  $\hat{\theta}_{n}^{\text{DR}}$, $\hat{\theta}_{n}^{\text{SL}}$ and $\hat{\theta}_{n}^{\text{SDR}}$. 
However, the DRCL has the weakest smoothness requirements on the loss function. Therefore, in this sense,  DRCL can be used more generally, especially when SL and sDRCL are not applicable, such as the quantile regression or the loss functions with the ReLU activation,  as demonstrated numerically in Section \ref{sec:simu-non-smooth}.  However, for the linear model, the asymptotic variances of the three estimators have more explicit expressions so that their efficiency can be directly compared as shown in the next proposition. 

\begin{proposition}\label{lemma:var} 
Consider the linear regression model $Y_{i} = X_{i}^{\top}\theta + \epsilon_{i}$ for $n$ i.i.d. copies of $(X_i, Y_i)$, where $\mathbb{E}(X_{i}) = 0$, $\mathbb{E}(X_{i}X_{i}^{\top}) = \Sigma_{x}$, $\mathbb{E}(\epsilon_{i}) = 0$ and $\mathrm{Var}(\epsilon_{i}) = \sigma^{2}$. 
Suppose that $\Sigma_{x}$ is invertible and $\{X_{i}\}_{i=1}^{n}$ are independent with $\{\epsilon_{i}\}_{i=1}^{n}$. 
For the DRDP mechanism in Algorithm \ref{alg:data_pre} with the DP parameters $\delta\in(0,1)$ and $\lambda>0$, we have 
\begin{longlist} 
    \item \label{var_0}
$\AVar(\hat{\theta}_{n}^{\text{SL}})=\Sigma_{x}^{-1}(\sigma^{2}\Sigma_{x}+\lambda^{2}\|\theta\|^{2}\Sigma_{x}+\sigma^{2}\lambda^{2}I_{d}+2\lambda^{4}\|\theta\|^{2}I_{d}+3\lambda^{4}\theta\theta^{\top})\Sigma_{x}^{-1}$,
    \item \label{var_1} 
    $\AVar(\hat{\theta}_{n}^{\text{SL}})-\AVar(\hat{\theta}_{n}^{\text{DR}})=(2-\delta^{-1})\Sigma_{x}^{-1}V\Sigma_{x}^{-1}$,
    \item \label{var_2}
    $\AVar(\hat{\theta}_{n}^{\text{SL}})-\AVar(\hat{\theta}_{n}^{\text{SDR}})=\delta\cdot\Sigma_{x}^{-1} V\Sigma_{x}^{-1}$, 
    \item \label{var_3}
    $\AVar(\hat{\theta}_{n}^{\text{DR}})-\AVar(\hat{\theta}_{n}^{\text{SDR}})=(\delta+\delta^{-1}-2)\Sigma_{x}^{-1}V\Sigma_{x}^{-1}$,
\end{longlist}
where 
$V = \lambda^{2}\|\theta\|^{2}\Sigma_{x}+\sigma^{2}\lambda^{2}I_{d}+2\lambda^{4}\|\theta\|^{2}I_{d}+(2+\delta)\lambda^{4}\theta\theta^{\top}$.
\end{proposition} 

\Cref{lemma:var} suggests that the sDRCL estimator is the most efficient estimator for the linear model over the entire range of $\delta \in (0,1)$, while the SL corrected-loss estimator is more efficient than the DRCL estimator for $\delta \in (0,1/2) $ and vice versa for $\delta \in (1/2,1)$.  
This proporsition also indicates that as $\lambda$ increases, the asymptotic variances of $\hat{\theta}_{n}^{\text{SL}}$, $\hat{\theta}_{n}^{\text{SDR}}$, and $\hat{\theta}_{n}^{\text{DR}}$ all increase. When $\delta$ decreases, the asymptotic variance of $\hat{\theta}_{n}^{\text{SL}}$ remains unchanged, while the asymptotic variance of $\hat{\theta}_{n}^{\text{DR}}$ increases and tends to infinity. The asymptotic variance of $\hat{\theta}_{n}^{\text{SDR}}$ also increases and tends to the asymptotic variance of $\hat{\theta}_{n}^{\text{SL}}$.

From the properties of the ZIL trade-off function in Section \ref{sec:privacy-guarantee}, it increases with the increase of $\lambda$ and decrease of $\delta$, meaning a higher level of privacy protection with the increase of the noise level. From \Cref{lemma:var}, for the linear model, the variances of the proposed estimators $\hat{\theta}_{n}^{\text{SDR}}$ and $\hat{\theta}_{n}^{\text{DR}}$ increase with the increase of $\lambda$ and decrease of $\delta$. This indicates that estimation efficiency is lower for a higher level of differential privacy, which is a price paid for protecting data confidentiality. However, this relationship may not generally hold, depending on the model and the estimator.
A counter-example can be made for the truncated mean estimator 
for data with added uniform noise; see the SM for details. }
%

%

\section{Simulation}
\label{sec:simulation}

In this section, we report results from simulation experiments designed to confirm the theoretical findings of the proposed DRCL and sDRCL estimators for DP M-estimation in the earlier sections under non-smooth $L_2$ loss, logistic regression, and quantile regression. To gain comparative insights, we also include the SL corrected-loss estimator $\hat{\theta}^{\text{SL}}_{n}$ for smooth loss functions and the approach of \cite{wang_correctedloss_2012} for quantile regression that smooths the check loss function, denoted as sCL. 

\subsection{Non-smooth loss}\label{sec:simu-non-smooth}  
To demonstrate the ability of the proposed DRCL procedure for M-estimation with non-smooth loss functions, we considered  three loss functions
\( \ell_{1}(x, \theta) = (\theta - \text{ReLU}(x))^{2} \), \( \ell_{2}(x, \theta) = (\theta - \mathbf{1}_{[0.5,1]}(x))^{2} \) and \( \ell_{3}(x, \theta) = (\theta - |\sin 2\pi x|)^{2} \) that are non-smooth with respect to \( x \), where \( \text{ReLU}(x) = \max(0, x) \) denotes the ReLU function and \( \mathbf{1}_{[0.5,1]}(x) \) denotes the indicator function on the interval $[0.5, 1]$.
The original data \( \{X_{i}\}_{i=1}^{n} \) were generated from the uniform distribution \( U(0, 1) \). Noisy data \( \{\tilde{X}_{i}^{(1)}\}_{i=1}^{n} \) and doubly randomized data \( \{\tilde{X}_{i}^{(2)}\}_{i=1}^{n} \) were generated using Algorithm \ref{alg:data_pre} with two sets of privacy parameters, \( \delta = 0.1, \lambda = 0.94 \) and \( \delta = 0.05, \lambda = 1.4 \). The true parameter \( \theta_{0,j} = \operatorname{argmin}_{\theta} \E \ell_{j}(X_{i}, \theta) \) for \( j = 1, 2, 3 \). As $X_i$ was \( U(0, 1) \),  \( \theta_{0,1} = \E \text{ReLU}(X_{i})=0.5 \), \( \theta_{0,2} = \E \mathbf{1}_{[0.5,1]}(X_{i})=0.5 \), and \( \theta_{0,3} = \E |\sin 2\pi X_{i}| =2/\pi\). The sample sizes were  \( n = 500, 1000, 2000, 3000, 4000\) and \(5000 \).  

\begin{figure}[ht]
    \centering
        \includegraphics[width=\linewidth]{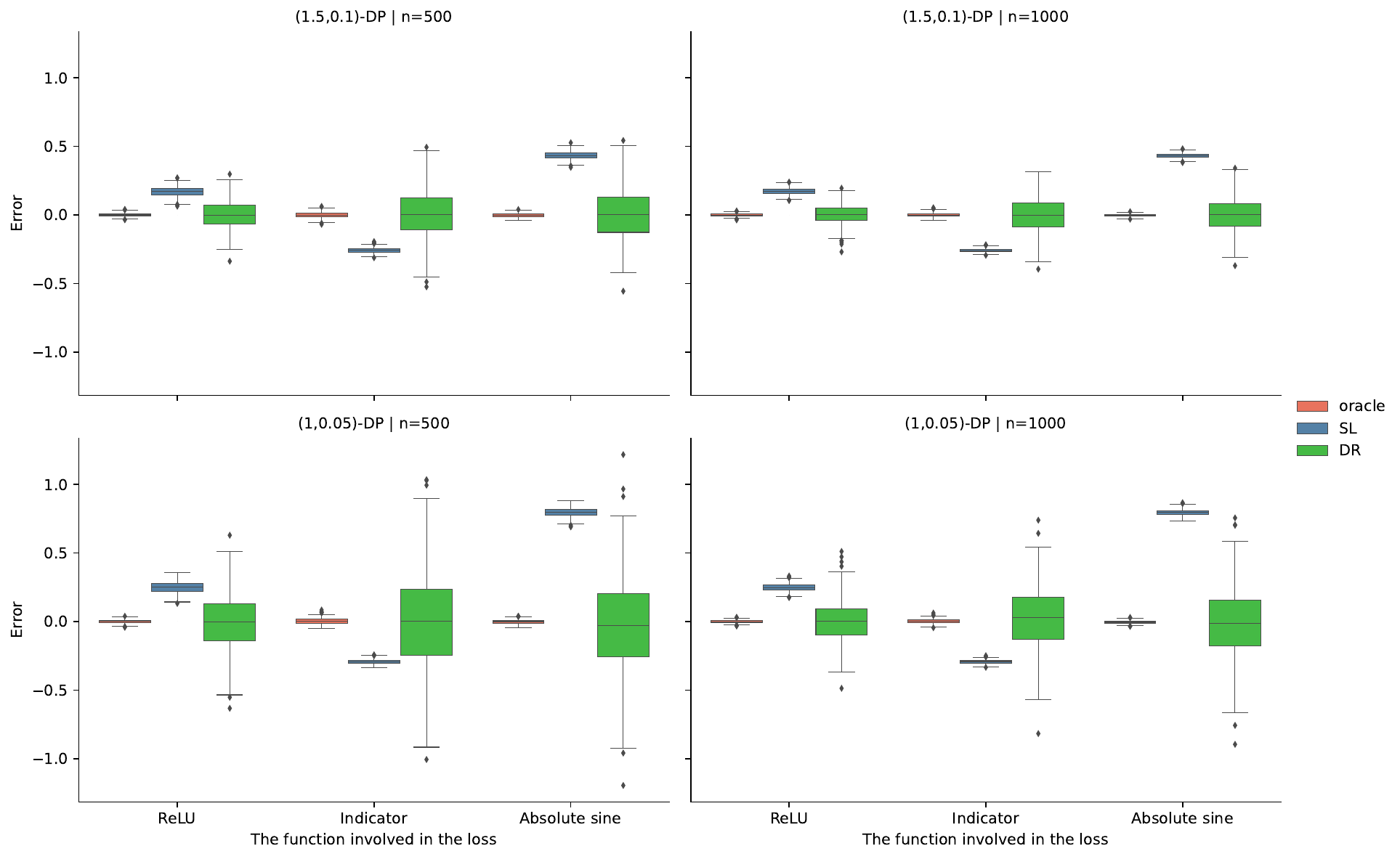}
    \caption{Box-plots of estimation errors of the oracle (left box), SL corrected-loss (middle box), and DRCL (right box) estimators for the ReLU, indicator, and absolute sine functions, respectively. The first row corresponds to \( \delta = 0.1, \lambda = 0.94 \) at  \( (1.5, 0.1) \)-DP, and the second row corresponds to \( \delta = 0.05, \lambda = 1.4 \) at \( (1, 0.05) \)-DP.}
    \label{fig:non-smooth}
\end{figure}

For each loss function, we performed M-estimation using three methods, the oracle estimator \( \hat{\theta}_{n,j}^{\text{ORA}} \) using the original data, the SL corrected-loss estimator \( \hat{\theta}_{n,j}^{\text{SL}}\) using $\{\tilde{X}_{i}^{(2)}\}$, and the DRCL estimator \( \hat{\theta}_{n,j}^{\text{DR}} \). Note that the SL estimator is designed for smooth loss functions which is unsuitable for those three non-smooth loss functions. 
Figure \ref{fig:non-smooth} displays the box plots of the estimation errors of the three DP M-estimators based on $5000$ simulations with the average root mean square errors shown in Table \ref{table:rmse_nonsmooth}. We only report the results for the sample sizes $n=500$ and $n=1000$, while results for other sample sizes are available in Table \ref{table:rmse_validate_nonsmooth} in the SM. The figure and the tables show that for all three loss functions, the SL estimator was not a consistent estimator of the true parameter. In contrast, the DRCL estimator was unbiased, and its standard error decreased with the increase of the sample size, which indicates its consistency and confirms the finding in \Cref{thm:identification}.

Under the first DP parameter setting of \( \delta = 0.1 \) and \( \lambda = 0.94 \), both \( \hat{\theta}_{n,j}^{\text{DR}} \) and \( \hat{\theta}_{n,j}^{\text{SL}} \) achieved \( (1.5, 0.1) \)-DP. Under the second DP setting of \( \delta = 0.05 \) and \( \lambda = 1.4 \), they achieved \( (1, 0.05) \)-DP. The $(1, 0.05)$-DP provides higher privacy protection than the $(1.5, 0.1)$-DP. 
As shown in Section \ref{sec:privacy-guarantee}, a larger $\lambda$ or smaller $\delta$ indicates higher privacy protection. Note that the variance of the ZIL noise $\text{ZIL}(\delta,\lambda^{2}I_{d})$ used in the Algorithm \ref{alg:data_pre} is $(1-\delta)\lambda^{2}I_{d}$. In general, the added noises with higher variance result in stronger differential privacy guarantees. 
It is observed that while the level of privacy protection increases, the performance of the SL and DRCL estimators deteriorates at the same sample size. Meanwhile, the variance of the DRCL estimator was larger than that of the oracle estimator, representing a cost of privacy protection, as indicated by Corollary \ref{thm:efficiency}.

\begin{table}[h]
\caption{The average root mean square errors of the oracle, SL corrected-loss (SL), and DRCL (DR) estimators for the parameters of the three non-smooth loss functions based on 5000 repetitions at $(1.5,0.1)$ and $(1.5,0.05)$-DP. The oracle estimator was calculated using the original non-corrupted data, which didn't protect data privacy.} 
\label{table:rmse_nonsmooth}
\begin{tabular}{cccccc}
\hline
n                     & DP                            & method & ReLU  & Indicator & Absolute sine \\ \hline
\multirow{6}{*}{500}  & None                          & Oracle & 0.012 & 0.023     & 0.015         \\
                      & \multirow{2}{*}{(1.5,0.1)-DP} & SL     & 0.173 & 0.259     & 0.433         \\
                      &                               & DR   & 0.105 & 0.183     & 0.170         \\ \cline{2-6} 
                      & None                          & Oracle & 0.013 & 0.022     & 0.013         \\
                      & \multirow{2}{*}{(1,0.05)-DP}  & SL     & 0.252 & 0.294     & 0.796         \\
                      &                               & DR   & 0.184 & 0.326     & 0.358         \\ \hline
\multirow{6}{*}{1000} & None                          & Oracle & 0.009 & 0.016     & 0.011         \\
                      & \multirow{2}{*}{(1.5,0.1)-DP} & SL     & 0.174 & 0.258     & 0.433         \\
                      &                               & DR   & 0.072 & 0.128     & 0.123         \\ \cline{2-6} 
                      & None                          & Oracle & 0.009 & 0.016     & 0.010         \\
                      & \multirow{2}{*}{(1,0.05)-DP}  & SL     & 0.250 & 0.294     & 0.796         \\
                      &                               & DR   & 0.131 & 0.230     & 0.257         \\ \hline
\end{tabular}
\end{table}


\subsection{Logistic regression} 
We considered the logistic regression model 
$\Prob(Y_i = 1 | X_i, \beta) = \{1 + \exp(-X_i^{\top} \beta)\}^{-1}$, where $\beta = (\beta_1, \ldots, \beta_6)^{\top}$, the covariates $\{X_i\}_{i=1}^{n}$ were i.i.d.\ generated from a truncated multivariate normal distribution \( \mathcal{N}(0_6, I_6)\) over a rectangle formed by a lower bound \( - 1_6 \) and an upper bound \( 1_6 \), where $0_p$ and $1_p$ denote the $p$-dimensional vectors of 0 and 1, respectively.
The noisy covariates \( \{\tilde{X}_{i}^{(1)}\}_{i=1}^{n} \) and the doubly randomized covariates \( \{\tilde{X}_{i}^{(2)}\}_{i=1}^{n} \) were generated using Algorithm \ref{alg:data_pre}, with privacy parameters \( \lambda \) and \( \delta \) set to \( (0.5, 0.2) \) and \( (1, 0.2) \), respectively. These settings provided privacy guarantees of \( T_{6,4,0.2} \)-ADP and \( T_{6,2,0.2} \)-ADP for the covariates, respectively. 
The true parameter  \( \beta_{\ast} = 1_6 \). The sample sizes considered were \( n = 5000, 7500\) and \(10000 \). 

We compare five M-estimators. The first method used the original data and minimized 
\(
\sum_{i=1}^{n} \{(1 - Y_i) \cdot X_i^{\top} \beta + \log(1 + \exp(-X_i^{\top} \beta))\}
\)
to obtain the oracle estimator \( \hat{\beta}_n^{\text{ORA}} \). The second one directly minimized the corrupted loss function
\(
\sum_{i=1}^{n} \{(1 - Y_i) \cdot \tilde{X}_i^{(1)\top} \beta + \log(1 + \exp(-\tilde{X}_i^{(1)\top} \beta))\}
\)
to obtain the naive estimator \( \hat{\beta}_n^{\text{NAI}} \). The other three methods are the SL corrected-loss (SL), DRCL (DR) and sDRCL (sDR) estimators.

The RMSEs of the estimated parameters based on 5000 replications are presented in Table \ref{table:logit_reg}. From this table, the naive estimator which did not take any measure to counter the added noises had the worse RMSEs in all cases, which was expected. Among the three differential private estimators (SL, sDRCL, and DRCL), the RMSE decreased as the sample size increased. However, as the privacy protection level was increased from \( T_{6,4,0.2} \)-ADP to \( T_{6,2,0.2} \)-ADP, the RMSE of the three estimators increased. Notably, the RMSEs of the SL, sDRCL, and DRCL estimators were always higher than that of the oracle estimator, reflecting the cost of ensuring privacy protection. 
Meanwhile, the sDRCL estimator consistently yielded a smaller RMSE compared to the SL estimator, and the DRCL estimator had a larger RMSE than both sDRCL and SL. This was consistent with the theoretical conclusion about the efficiency of the three estimators obtained under the linear regression setting in \Cref{lemma:var}.

\begin{table}[h]
\caption{The average root mean square errors of the oracle, naive, SL corrected-loss (SL), sDRCL (sDR) and DRCL (DR) estimators for the logistic regression coefficients based on 5000 simulations at $T_{6,4,0.2}$-DP and $T_{6,2,0.2}$-DP. The oracle estimator was calculated using the original non-corrupted data, which didn’t protect data privacy.}
\label{table:logit_reg}
\begin{tabular}{ccccccccc}
\hline
n                       & attribute-level DP        & method & $\beta_{1}$ & $\beta_{2}$ & $\beta_{3}$ & $\beta_{4}$ & $\beta_{5}$ & $\beta_{6}$ \\ \hline
\multirow{10}{*}{5000}  & None                      & Oracle & 0.105       & 0.107       & 0.106       & 0.107       & 0.103       & 0.105       \\
                        & \multirow{4}{*}{$T_{6,4,0.2}$}  & Naive  & 0.728       & 0.730       & 0.731       & 0.728       & 0.730       & 0.731       \\
                        &                           & SL     & 0.270       & 0.265       & 0.262       & 0.267       & 0.270       & 0.271       \\
                        &                           & sDR  & 0.244       & 0.239       & 0.234       & 0.238       & 0.242       & 0.242       \\
                        &                           & DR   & 0.495       & 0.498       & 0.495       & 0.489       & 0.494       & 0.495       \\ \cline{2-9} 
                        & None                      & Oracle & 0.104       & 0.103       & 0.107       & 0.111       & 0.098       & 0.102       \\
                        & \multirow{4}{*}{$T_{6,2,0.2}$} & Naive  & 0.905       & 0.901       & 0.913       & 0.907       & 0.907       & 0.911       \\
                        &                           & SL     & 0.610       & 0.618       & 0.586       & 0.600       & 0.609       & 0.622       \\
                        &                           & sDR  & 0.536       & 0.542       & 0.517       & 0.535       & 0.551       & 0.557       \\
                        &                           & DR   & 0.769       & 0.751       & 0.749       & 0.752       & 0.782       & 0.766       \\ \hline
\multirow{10}{*}{7500}  & None                      & Oracle & 0.086       & 0.087       & 0.085       & 0.086       & 0.088       & 0.086       \\
                        & \multirow{4}{*}{$T_{6,4,0.2}$}  & Naive  & 0.727       & 0.729       & 0.728       & 0.728       & 0.730       & 0.730       \\
                        &                           & SL     & 0.217       & 0.218       & 0.215       & 0.216       & 0.218       & 0.217       \\
                        &                           & sDR  & 0.195       & 0.197       & 0.191       & 0.193       & 0.197       & 0.193       \\
                        &                           & DR   & 0.409       & 0.407       & 0.402       & 0.407       & 0.411       & 0.408       \\ \cline{2-9} 
                        & None                      & Oracle & 0.086       & 0.087       & 0.085       & 0.086       & 0.088       & 0.086       \\
                        & \multirow{4}{*}{$T_{6,2,0.2}$} & Naive  & 0.910       & 0.912       & 0.911       & 0.910       & 0.912       & 0.913       \\
                        &                           & SL     & 0.522       & 0.528       & 0.518       & 0.518       & 0.518       & 0.516       \\
                        &                           & sDR  & 0.445       & 0.455       & 0.437       & 0.441       & 0.447       & 0.438       \\
                        &                           & DR   & 0.706       & 0.705       & 0.707       & 0.713       & 0.713       & 0.713       \\ \hline
\multirow{10}{*}{10000} & None                      & Oracle & 0.074       & 0.075       & 0.075       & 0.073       & 0.075       & 0.076       \\
                        & \multirow{4}{*}{$T_{6,4,0.2}$}  & Naive  & 0.728       & 0.729       & 0.728       & 0.729       & 0.729       & 0.728       \\
                        &                           & SL     & 0.190       & 0.189       & 0.184       & 0.187       & 0.187       & 0.186       \\
                        &                           & sDR  & 0.170       & 0.168       & 0.165       & 0.168       & 0.169       & 0.168       \\
                        &                           & DR   & 0.355       & 0.348       & 0.351       & 0.353       & 0.356       & 0.360       \\ \cline{2-9} 
                        & None                      & Oracle & 0.074       & 0.075       & 0.075       & 0.073       & 0.075       & 0.076       \\
                        & \multirow{4}{*}{$T_{6,2,0.2}$} & Naive  & 0.913       & 0.914       & 0.913       & 0.914       & 0.913       & 0.913       \\
                        &                           & SL     & 0.460       & 0.461       & 0.452       & 0.459       & 0.461       & 0.458       \\
                        &                           & sDR  & 0.390       & 0.387       & 0.380       & 0.386       & 0.388       & 0.388       \\
                        &                           & DR   & 0.672       & 0.660       & 0.669       & 0.665       & 0.664       & 0.671       \\ \hline
\end{tabular}
\end{table}

\subsection{Quantile regression} 
We considered the median regression \( \text{Median}(Y_{i} \mid X_{i}) = \beta_{0} + X_{i}^{\top} \beta \) as in \cite{pan_note_2022}, where $\beta = (\beta_1, \ldots, \beta_6)^{\top}$.  
The covariates \( \{ X_i = (X_{i1}, \dots, X_{i6})^{\top}\}_{i=1}^{n} \) are i.i.d.\ generated from a multivariate truncated normal distribution \( \mathcal{N}(0_6, I_6)\) with the lower and upper truncation bounds \( - 1_6 \) and \( 1_6 \), respectively. Let \( \{\varepsilon_{i}\}_{i=1}^{n} \) be i.i.d. \( N(0, \sigma^{2}) \) with $\sigma=1$ and the true parameters $\beta_{\ast, 0} = 1$ and $\beta_{\ast} = 1_6$, so that \( 1 + \sum_{j = 1}^{6} X_{ij} \) is the conditional median of \( Y_{i} \) for \( i = 1, \dots, n \). 
The noisy covariates \( \{\tilde{X}_{i}^{(1)}\}_{i=1}^{n} \) and the doubly randomized covariates \( \{\tilde{X}_{i}^{(2)}\}_{i=1}^{n} \) were generated using Algorithm \ref{alg:data_pre} with the privacy parameters \( \lambda \) and \( \delta \) set to \( (2, 0.2) \) and \( (2.5, 0.2) \), respectively. These settings provided \( T_{6,2,0.2} \)-ADP and \( T_{6,0.8,0.2} \)-ADP for the covariates.
The sample sizes were \( n = 2500, 5000\), and \(7500 \).

We compared four methods for estimation. The first method used original data to minimize \( \sum_{i=1}^{n} |Y_{i} - X_{i}^{\top} \beta| \) and obtain the oracle estimator \( \hat{\beta}_{n}^{\text{ORA}} \). The second method directly minimized \( \sum_{i=1}^{n} |Y_{i} - \tilde{X}_{i}^{(1)\top} \beta| \) to obtain the naive estimator \( \hat{\beta}_{n}^{\text{NAI}} \). The third method, proposed by \cite{wang_correctedloss_2012}, minimized a smoothed corrected loss by kernel smoothing the absolute value function, which is denoted as sCL. The fourth one was the proposed DRCL method. We did not consider the SL corrected-loss and sDRCL methods as the check loss is not twice differentiable everywhere. The average RMSEs of the parameter estimates based on 5000 replications are presented in Table \ref{table:quantile_reg}. 

It is observed from Table \ref{table:quantile_reg} that the DRCL estimator had a smaller RMSE than that of sCL. The RMSE of the DRCL estimator decreases as the sample size increases. However, as the differential privacy (DP) level improves, for instance, changing from \( T_{6,2,0.2} \)-ADP to \( T_{6,0.8,0.2} \)-ADP, the RMSE of the DRCL estimator increases.
Except for the oracle estimator, the other three estimators were differentially private. 
The RMSE of the DRCL estimator is always larger than that of the oracle estimator, as indicated by Corollary \ref{thm:efficiency}.

\begin{table}[h]
\caption{The average root mean square errors of the oracle, naive, sCL \citep{wang_correctedloss_2012} and the proposed DRCL (DR) estimators for the median regression coefficients based on $5000$ simulations at $T_{6,2,0.2}$-DP and $T_{6,0.8,0.2}$-DP. The oracle estimator was calculated using the original non-corrupted data, which didn’t protect data privacy.
}
\label{table:quantile_reg}
\begin{tabular}{cccccccccc}
\hline
n                     & attribute-level DP        & method             & $\beta_{0}$ & $\beta_{1}$ & $\beta_{2}$ & $\beta_{3}$ & $\beta_{4}$ & $\beta_{5}$ & $\beta_{6}$ \\ \hline
\multirow{8}{*}{2500} & None                      & oracle             & 0.020       & 0.037       & 0.038       & 0.038       & 0.037       & 0.037       & 0.037       \\
                      & \multirow{3}{*}{$T_{6,2,0.2}$}  & naive              & 0.037       & 0.912       & 0.912       & 0.911       & 0.911       & 0.912       & 0.912       \\
                      &                           & sCL & 0.111       & 0.632       & 0.630       & 0.637       & 0.631       & 0.634       & 0.635       \\
                      &                           & DR               & 0.094       & 0.443       & 0.438       & 0.444       & 0.438       & 0.446       & 0.439       \\ \cline{2-10} 
                      & None                      & oracle             & 0.020       & 0.037       & 0.038       & 0.038       & 0.037       & 0.037       & 0.037       \\
                      & \multirow{3}{*}{$T_{6,0.8,0.2}$} & naive              & 0.037       & 0.943       & 0.942       & 0.942       & 0.942       & 0.943       & 0.942       \\
                      &                           & sCL & 0.143       & 0.738       & 0.731       & 0.734       & 0.733       & 0.737       & 0.735       \\
                      &                           & DR               & 0.100       & 0.499       & 0.503       & 0.503       & 0.512       & 0.507       & 0.504       \\ \hline
\multirow{8}{*}{5000} & None                      & oracle             & 0.014       & 0.026       & 0.026       & 0.027       & 0.026       & 0.027       & 0.026       \\
                      & \multirow{3}{*}{$T_{6,2,0.2}$}  & naive              & 0.027       & 0.911       & 0.911       & 0.911       & 0.912       & 0.911       & 0.911       \\
                      &                           & sCL & 0.074       & 0.484       & 0.488       & 0.487       & 0.487       & 0.491       & 0.488       \\
                      &                           & DR               & 0.061       & 0.302       & 0.296       & 0.299       & 0.296       & 0.300       & 0.297       \\ \cline{2-10} 
                      & None                      & oracle             & 0.014       & 0.026       & 0.026       & 0.027       & 0.026       & 0.027       & 0.026       \\
                      & \multirow{3}{*}{$T_{6,0.8,0.2}$} & naive              & 0.028       & 0.942       & 0.942       & 0.942       & 0.942       & 0.942       & 0.942       \\
                      &                           & sCL & 0.101       & 0.636       & 0.638       & 0.636       & 0.640       & 0.638       & 0.637       \\
                      &                           & DR               & 0.065       & 0.375       & 0.376       & 0.377       & 0.374       & 0.380       & 0.375       \\ \hline
\multirow{8}{*}{7500} & None                      & oracle             & 0.011       & 0.021       & 0.022       & 0.021       & 0.022       & 0.021       & 0.022       \\
                      & \multirow{3}{*}{$T_{6,2,0.2}$}  & naive              & 0.021       & 0.911       & 0.912       & 0.911       & 0.912       & 0.912       & 0.912       \\
                      &                           & sCL & 0.058       & 0.392       & 0.395       & 0.391       & 0.392       & 0.392       & 0.394       \\
                      &                           & DR               & 0.049       & 0.246       & 0.245       & 0.244       & 0.244       & 0.242       & 0.240       \\ \cline{2-10} 
                      & None                      & oracle             & 0.011       & 0.021       & 0.022       & 0.021       & 0.022       & 0.021       & 0.022       \\
                      & \multirow{3}{*}{$T_{6,0.8,0.2}$} & naive              & 0.022       & 0.942       & 0.942       & 0.942       & 0.942       & 0.942       & 0.942       \\
                      &                           & sCL & 0.081       & 0.556       & 0.563       & 0.558       & 0.559       & 0.558       & 0.560       \\
                      &                           & DR               & 0.052       & 0.301       & 0.304       & 0.300       & 0.300       & 0.303       & 0.296       \\ \hline
\end{tabular}
\end{table}


\section{Discussion}\label{sec:discuss}

We have developed a versatile DP mechanism and its estimation procedure. The  'versatility' 
means that the DP procedure applies to {\it general} M-estimation tasks under minimum conditions on the loss function, is {\it multitasking} in that it allows different tasks from an unlimited number of data analysts, and is {\it online-adaptive} to increasing data volume.
This paper has three significant contributions. First, the proposed ZIL and DRDP privacy mechanisms are versatile. In contrast, some existing well-known methods like the noisy-SGD and the Perturbed-Histogram are not versatile. 
Second, the trade-off function and the privacy protection level of the ZIL mechanism are derived and carefully studied. Third, a new method to recover the target loss function is proposed to consistently estimate the underlying parameters, which works for non-smooth loss functions and is easy to implement without numerical integration and differentiation.

In contrast, the noisy-SGD and Perturbed-Histogram are not as versatile as the proposed procedure. 
This is because, in real applications, the data analysts in their various data analytic tasks, have to request the mechanism to repeatedly execute the noisy-SGD or Perturbed-Histogram procedures for hyper-parameter selection, new data arrival, or different analysis tasks, where each execution would consume a certain amount of privacy budget. 
Specifically, for different models or estimation tasks, the noisy-SGD needs to be rerun repeatedly, and the hyperparameters, such as epochs, learning rate, and batch size, need to be reselected each time. 
For the Perturbed-Histogram, 
the arrival of new data requires recalculation of the histogram over the entire dataset, which consumes the privacy budget allocated to the original dataset.
Therefore, when applying noisy-SGD or Perturbed-Histogram under limited privacy budgets, analysts face restrictions, making the approaches not versatile. 
In contrast, the proposed DP mechanism and the associated DRCL estimation procedure provide a solution for this problem. %

\begin{acks}[Acknowledgments]
\end{acks}

\begin{funding}

\end{funding}

\begin{supplement}
\stitle{Supplement to "Versatile differentially private learning for general loss functions"}
\sdescription{In the supplementary material, we present technical details, proofs of main theorems and additional numerical results.}
\end{supplement}

\bibliographystyle{imsart-nameyear} 
\bibliography{Ref}       

\end{document}